\documentclass[11pt]{article}
\usepackage{mathrsfs}
\usepackage{natbib,graphicx,setspace,lscape,longtable}
\usepackage{mathrsfs,amsmath,amsthm,amssymb,color}
\usepackage{natbib,epsfig,graphicx,pdfpages}
\usepackage{rotating}
\usepackage{subfigure}
\usepackage{enumerate}
\usepackage{graphicx}
\usepackage{algorithmic}
\usepackage{algorithm}

\usepackage{url}
\usepackage{hyperref}
\usepackage[table]{xcolor}
\usepackage{colortbl}
\bibpunct{(}{)}{;}{a}{,}{,}

\newtheorem{theorem}{Theorem}
\newtheorem{lemma}{Lemma}
\newtheorem{proposition}{Proposition}
\newtheorem{remark}{Remark}
\usepackage{amsthm}
\newtheorem{corollary}{Corollary}

\newtheorem{assumption}{{Assumption}}
\def\beq{\begin{equation}}
\def\eeq{\end{equation}}
\def\beqr{\begin{eqnarray}}
\def\eeqr{\end{eqnarray}}
\def\beqrs{\begin{eqnarray*}}
\def\eeqrs{\end{eqnarray*}}
\def\bet{\begin{theorem}}
\def\eet{\end{theorem}}
\def\bel{\begin{lemma}}
\def\eel{\end{lemma}}
\def\bep{\begin{proposition}}
\def\eep{\end{proposition}}
\def\bg{\begin{figure}[tbph]\begin{center}}
\def\eg{\end{center}\end{figure}}

\def\bc{\begin{center}}
\def\ec{\end{center}}

\def\wt{\widetilde}
\def\wh{\widehat}

\def\diag{\mbox{diag}}

\numberwithin{equation}{section}

\newcommand{\Cov}{\textnormal{Cov}}

\newcommand{\bA}{{\mathbf A}}
\newcommand{\bB}{{\mathbf B}}
\newcommand{\bF}{{\mathbf F}}

\newcommand{\bG}{{\mathbf G}}
\newcommand{\bH}{{\mathbf H}}
\newcommand{\bI}{{\mathbf I}}
\newcommand{\bK}{{\mathbf K}}

\newcommand{\bM}{{\mathbf M}}

\newcommand{\bQ}{{\mathbf Q}}
\newcommand{\bP}{{\mathbf P}}

\newcommand{\bS}{{\mathbf S}}

\newcommand{\bU}{{\mathbf U}}
\newcommand{\bV}{{\mathbf V}}
\newcommand{\bW}{{\mathbf W}}

\newcommand{\bY}{{\mathbf Y}}

\newcommand{\ba}{{\mathbf a}}
\newcommand{\bb}{{\mathbf b}}

 \newcommand{\bfc}{{\mathbf c}}
\newcommand{\be}{{\mathbf e}}
\newcommand{\bff}{{\mathbf f}}
 \newcommand{\bgg}{{\mathbf g}}

\newcommand{\bt}{{\mathbf t}}
\newcommand{\bs}{{\mathbf s}}
\newcommand{\bu}{{\mathbf u}}
\newcommand{\bv}{{\mathbf v}}
\newcommand{\bw}{{\mathbf w}}
\newcommand{\bx}{{\mathbf x}}
\newcommand{\by}{{\mathbf y}}

\newcommand{\bbeta}  {\boldsymbol{\beta}}
\newcommand{\bfeta}  {\boldsymbol{\eta}}

\newcommand{\bOmega}{\boldsymbol{\Omega}}

\newcommand{\bSigma}{\boldsymbol{\Sigma}}

\newcommand{\bgamma}{\boldsymbol{\gamma}}

\newcommand{\bTheta} {\boldsymbol{\Theta}}
\newcommand{\bPhi} {\boldsymbol{\Phi}}

\newcommand{\bC}{{\mathbf C}}
\newcommand{\bD}{{\mathbf D}}

\newcommand{\ve}{{\varepsilon}}
\renewcommand{\epsilon}{{\ve}}
\renewcommand{\hat}{\widehat}

\def\wt{\widetilde}

\newcommand{\tr}{\mbox{tr}}
\def\JRSSB{{\sl Journal of the Royal Statistical Society}, {\bf B}}

\def\JASA{{\sl Journal of the American Statistical Association}}

\begin{document}

\title{Divide-and-Conquer: 
A Distributed Hierarchical Factor Approach to Modeling Large-Scale Time Series Data}
\author{
Zhaoxing Gao$^1$ and Ruey S. Tsay$^2$ \\
$^1$Department of Mathematics, Lehigh University\\
$^2$Booth School of Business, University of Chicago
}

 \date{}

\maketitle

\begin{abstract}
This paper proposes a hierarchical approximate-factor approach to analyzing high-dimensional, 
large-scale heterogeneous time series data using distributed computing. The new method  employs a multiple-fold dimension reduction procedure using Principal Component Analysis (PCA) and shows great promises for modeling large-scale data that  cannot be stored nor analyzed by a single machine. 
Each computer  at the basic level performs a PCA to extract common 
factors among  the time series assigned to  it and transfers those factors to one and only 
one node of the second level. Each 2nd-level computer collects the common factors from its subordinates and performs another PCA  
to select the 2nd-level common factors. This process is repeated until  the central server is reached, which collects 
common factors from its direct subordinates and performs a final PCA to 
select the global common factors. The noise terms of the 2nd-level approximate factor model 
are the unique common factors of the 1st-level clusters.  
We focus on the case of 2 levels in our theoretical derivations, but the idea can easily be 
generalized to any finite number of hierarchies.   We discuss 
some clustering methods  when the group memberships are unknown 
and introduce a new diffusion index approach to forecasting. We further extend the analysis 
to unit-root nonstationary time series. Asymptotic properties of the proposed method are derived for the diverging  dimension of the data in each computing unit 
and the sample size $T$. We use both simulated data and real examples to assess the performance of the proposed method in finite samples, and  compare our method with the commonly used ones in the literature concerning the forecastability of extracted factors.
\end{abstract}

\noindent {\sl Keywords}: Distributed Factor Model, Large-scale Time Series, Principal Component Analysis, Commonality, Heterogeneity

\newpage

\section{Introduction}

Multivariate time series with low-dimensional factor structures have been widely used in 
many scientific fields, including economics,  finance, and statistics, among others. Asset returns 
in finance are often modeled as  functions of a small number of factors, see \cite{ross1976}, \cite{StockWatson_1989},  \cite{famafrench1993,famafrench2015}, and \cite{StockWatson_1998}. Macroeconomic variables of multiple countries are often found to have comovements, see \cite{GregoryHead1999} and \cite{forni2000}. For large-scale data,  factor models not only provide an effective way to extract useful common information, but also reveal their  low-dimensional structure when the dimension is high.  Indeed, various factor models have been developed in the literature for multivariate and, recently, high-dimensional time series. See the approximate factor models in \cite{chamberlain1983}, \cite{StockWatson_1989,StockWatson_2002a,StockWatson_2002b}, \cite{BaiNg_Econometrica_2002}, and \cite{Bai_Econometrica_2003}; the dynamic factor models in \cite{forni2000,forni2005}; the factor models with dynamically dependent factors in \cite{lamyao2012} and \cite{gaotsay2018a,gaotsay2020a, gaotsay2020b}, among others. Other dimension reduction techniques for time series with a low dimensional structure are also available, 
see \cite{BoxTiao_1977}, \cite{TiaoTsay_1989}, \cite{Tsay_2014}, and the references therein. However, all available methods require the accessibility to the entire data set of interest by 
a single computer, and such a requirement has become increasingly infeasible with the rapid developments in information science and technology.

A fundamental challenge in modeling large-scale, high-dimensional time series is the storage of 
big data with huge sample size along the time horizon and a large number of series from the spatial domain. In many applications, we need to analyze internet data containing billions, or even trillions, of data points across spatial dimension and time horizon, making even a linear pass of the whole data set infeasible. In addition, such data are often highly dynamically dependent and 
may not be stored in a centralized machine. For example, stock returns of different exchanges and different countries may be stored in different databases, macroeconomic data  are often collected and stored in different government agencies, and the 
high-frequency spatio-temporal data, such as the real-time PM$_{2.5}$ indexes, are usually monitored and collected by local agencies across a country. Consequently, it has become  increasingly difficult to analyze them due to the communication cost, privacy concern, and data security, among others. Therefore, a distributed statistical method and computing architecture is needed. The fundamental approach to store and process such data is to divide and conquer with the basic idea being to partition a large problem into many tractable sub-problems. The sub-problems are tackled in parallel by different processing units. Intermediate results from individual machines are then combined to be further processed on the central server to yield the final output. Many distributed  methods following this approach have been developed during the past decades. See the distributed regression methods in \cite{zhang2013} and \cite{chenxie2014}, the distributed PCA in \cite{qu2002} and \cite{fan2019} with a horizontal partition regime, where each server contains a small portion of the data of all subjects,  
and  another distributed PCA in \cite{kargupta2001}, \cite{lietal2011}, and 
\cite{BertrandMoonen2014} with a vertical partition regime, where each server contains all the data of a small subset of the subjects. However, most of the methods mentioned above were developed for independent and identically distributed data, and they cannot tackle high-dimensional  time series directly.  In addition, the statistical guarantees are 
absent in some of the aforementioned literature.

\begin{figure}
\begin{center}
{\includegraphics[width=0.95\textwidth]{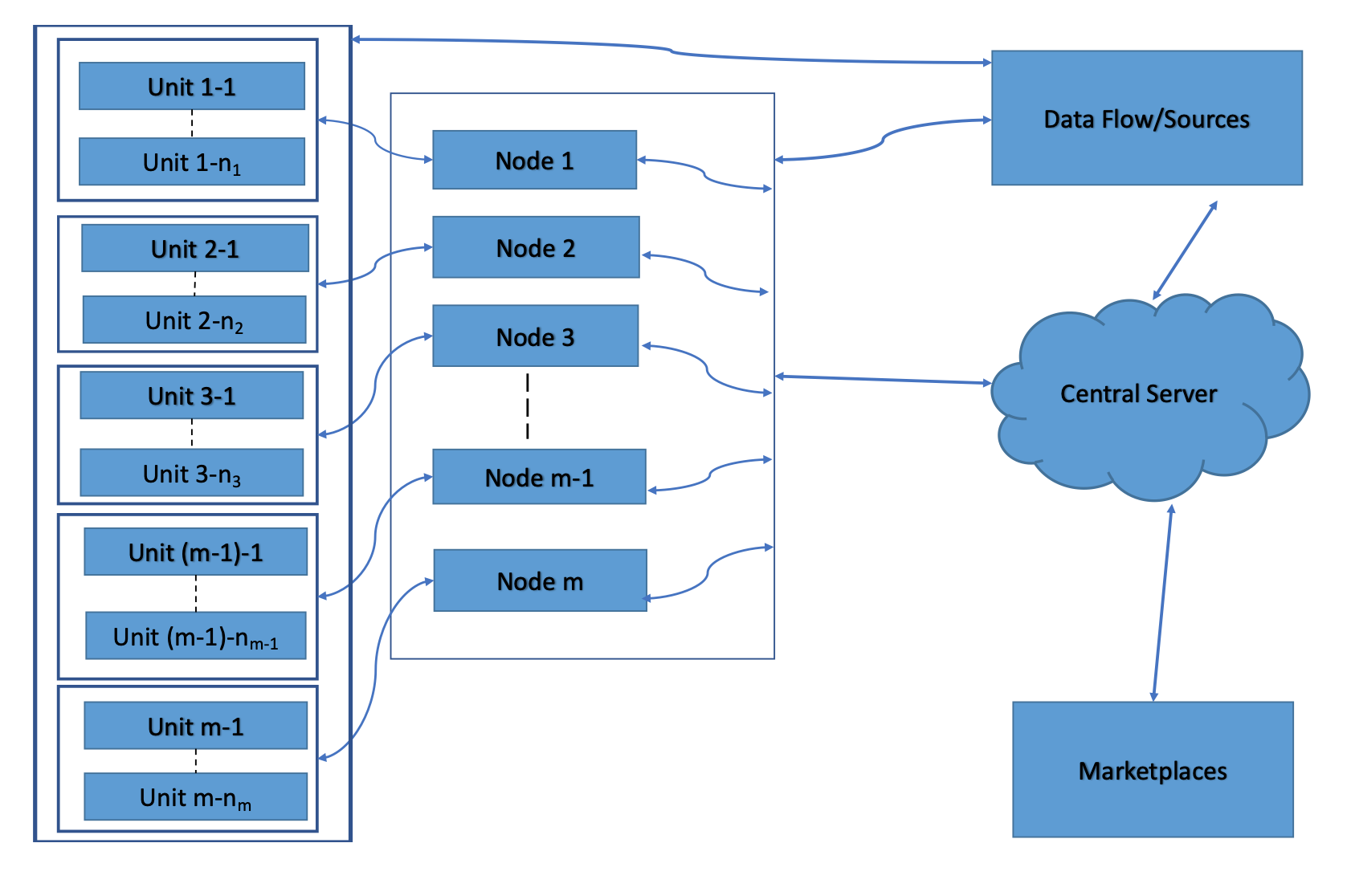}}
\caption{A 3-layer sketch of distributed computing framework for the proposed hierarchical factor modeling of large-scale time series data.}\label{fig0}
\end{center}
\end{figure}

The goal of this paper is to propose a hierarchical factor  framework for modeling large-scale time series data. When data are stored across multiple machines, or data centers, the traditional factor modeling requires analyzing the whole data set in a central server, resulting in excess computational cost caused by data transfer and communication.  Our proposed framework can 
overcome this difficulty efficiently. 
To illustrate, we sketch a 3-layer hierarchical factor modeling framework in Figure~\ref{fig0}, where 
the first layer has $n_1+\cdots+n_m$ processing units with $n_i$ of them connected to the 
$i$-th node in the second layer, and the $m$ nodes of the second layer are connected to the central 
server, which forms the 3rd layer. 
Suppose the marketplaces submit orders to the central server and expect to receive its responses later. 
Here the relevant data are either stored in the first-layer units, or assigned to them by the central server based on the capacities of the units and/or the problem under study. 
Each 1st-layer unit performs a PCA to extract common factors from the data assigned to it 
and sends those common factors to its connected node in the 2nd layer. After collecting the common factors sent to it, each node of the 2nd layer 
performs, again, a PCA to obtain the 2nd-level common factors. 
It then transfers those extracted common factors  to the central server, 
and the central server performs yet another PCA to extract the global common factors using the 
2nd-level common factors it received. 
Clearly, the idea of hierarchical factor analysis can easily be extended to more layers. In this way, 
the proposed scheme is similar to that of deep neural networks (DNN) with multiple layers, where each computing unit or node can be treated as a neuron. However, our objective is different from 
that of DNN, because here each neuron only computes linear combinations of the  factors it receives from its direct subordinates via PCA.

We focus on the scenario of 2 levels in our methodological formulations and theoretical derivations because it serves as a building-block for the general case, and the idea can easily be 
generalized to any finite number of hierarchies.  Specifically, we consider the $m$ computing nodes and the central server of Figure ~\ref{fig0}, in which each node contains all observations of 
a small subset of the $p$-dimensional time series. For a  $p$-dimensional time series 
and $m$ servers, 
we first apply PCA to the $p_i$-dimensional time series stored on the $i$-th node to 
obtain some common factors for the $i$-th group (or cluster), 
which are called {\em within group factors}. 
The central server then collects all within group factors to form a new vector time series, and  performs another PCA to select higher-level common factors. 
The resulting common factors serve as the global factors (or {\em between group factors}) 
that capture the cross-sectional information across all the groups, and the idiosyncratic terms 
of the 2nd-level PCA would be the {\em group-specific factors} of the corresponding 1st-level group.  This approach involves a two-fold dimension reduction procedure using PCA, and is in line with the distributed computing with a vertical partition regime. 
Thus, it is different from the one in \cite{fan2019}, 
and it is also different from the one in \cite{kargupta2001}, which only focuses on a 
single-stage PCA to reduce the dimension of each group without providing rigorous statistical guarantees. Asymptotic properties of the proposed method are established for diverging group dimension $p_i$ and the sample size $T$. We use both simulated data and real examples to assess the performance of the proposed method in finite samples. In the presence of group structure and 
assuming that the data set can be analyzed by a single machine, we apply the ratio-based method in \cite{lamyao2012} and \cite{ahn2013} and the BIC-type criterion in \cite{BaiNg_Econometrica_2002} to the whole data set to select the number of common factors.  We find that the former tends to find the number of global factors, whereas the latter selects the total number of group specific factors, as the number of common factors of the whole data. We also compare our method with the commonly used ones in the literature concerning the forecastability of the extracted factors, and find that our proposed new diffusion-index model outperforms the traditional one in forecasting eight commonly 
employed U.S. economic time series.

From a statistical perspective, it is worth mentioning that our resulting model is similar to the dynamic hierarchical factor models in \cite{Moench2013}, the dynamic factor model in \cite{alonso2020} with cluster structure, and the panel data model in \cite{andobai2017} with predictors and factor structures. However, their methods can only handle stationary time series, and require storing the entire data on a single machine, resulting in high computational cost 
for large data sets. In addition,  the statistical guarantees remain absent in \cite{Moench2013} and \cite{alonso2020}.  Our proposed method is  
different from those of the aforementioned literature, and its statistical properties are 
established for both stationary and unit-root nonstatinary time series data under some reasonable assumptions. Furthermore, when the data are in small-scale and stored in a single machine, our method provides an alternative way to modeling panel time series data with global and group-specific factor structures if a clustering method is available. We discuss this issue in Section 2.3 and 
propose a variation-based clustering method to distribute the data to different nodes 
if the group memberships are not available. This method for distributed computing 
is of independent  interest by itself.

The contributions of this paper are multi-fold. First, the proposed distributed framework accommodates the large-scale features of modern big data and can also be used to analyze independent and identically distributed data with parallel computing. 
The procedure is easy to implement and addresses two important features of the data:  Heterogeneity and Commonality, where the heterogeneity  is preserved in the group specific factors of the first PCA, and  the commonality of different groups is accounted for by the global factors obtained in the subsequent analysis.  
Second, unlike most of the distributed 
and hierarchical modeling procedures available in the literature, we established statistical guarantees of the proposed method provided that the number of machines used is not extremely large. Third, the framework is extended to unit-root nonstationary time series to extract the common stochastic trends of the data.  We also briefly discuss ways to incorporate horizontal partition regime into the proposed 
vertical regime for analyzing extremely large-scale data.  
Finally, we propose a new diffusion-index model for time series forecasting. This new model makes use of the global common factors and the group specific factors as predictors, and it is able to embrace the empirical findings in \cite{BoivinNg_joe_2006} that factors extracted from a smaller number of series might yield 
satisfactory or even better forecasting results.


The rest of the paper is organized as follows. We introduce the proposed model and 
estimation methodology in Section \ref{sec2} and 
study the theoretical properties of the proposed model and its associated 
 estimates in Section \ref{sec3}. The proposed method is extended to unit-root nonstationary time series data in Section \ref{sec4}. Section \ref{sec5} briefly discusses ways to combine the horizontal and vertical partitions together when the sample size $T$ is extremely large.
Numerical studies with both simulated and real data sets are given in Section \ref{sec6}, and 
Section \ref{sec7} provides some concluding remarks. 
All technical proofs are given in an Appendix. Throughout the article,
 we use the following notation. For a $p\times 1$ vector
$\bu=(u_1,..., u_p)',$  $||\bu||_2 =\|\bu'\|_2= (\sum_{i=1}^{p} u_i^2)^{1/2} $
is the Euclidean norm, $\|\bu\|_\infty=\max_{1\leq i\leq p}|u_i|$ is the $\ell_\infty$-norm, and $\bI_p$ denotes a $p\times p$ identity matrix. For a matrix $\bH=(h_{ij})$, $\|\bH\|_1=\max_j\sum_i|h_{ij}|$, $|\bH|_\infty=\max_{i,j}|h_{ij}|$,  $\|\bH
\|_F=\sqrt{\sum_{i,j}h_{ij}^2}$ is the Frobenius norm, $\|\bH
\|_2=\sqrt{\lambda_{\max} (\bH' \bH ) }$ is the operator norm, where
$\lambda_{\max} (\cdot) $ denotes for the largest eigenvalue of a matrix, and $\|\bH\|_{\min}$ is the square root of the minimum non-zero eigenvalue of $\bH'\bH$. The superscript ${'}$ denotes 
the transpose of a vector or matrix. We also use the notation $a\asymp b$ to denote $a=O(b)$ and $b=O(a)$.

\section{Models and Methodology}\label{sec2}

\subsection{Setting}
Consider a 2-level framework. 
Let $\by_t= (\by_{1,t}',\by_{2,t}'...,\by_{m, t}')'$ be an observable $p$-dimensional  time series consisting of $m$ groups with $\by_{i,t}=(y_{i1,t},...,y_{i p_i, t})'\in \mathbb{R}^{p_i}$, for $1\leq i\leq m$, stored on $m$ machines or servers, and $p_1+p_2+...+p_m=p$.  We consider the 
case of large $p$, but slowly diverging $m$. If there are more than two layers, $\{\by_{i,t},t=1,...,T\}$ may consist of stacked common factors from its associated groups in the previous layer. For simplicity, we assume that $E(\by_{i,t})={\bf 0}$ and 
$\{\by_{i,t}| t=1,...,T\}$ admits a latent factor structure:
\begin{equation}\label{m-factor}
\by_{i,t}=\bA_i\bgg_{i,t}+\be_{i,t},\quad i=1,...,m,
\end{equation}
where $\bgg_{i,t}=(g_{i1,t},...,g_{ir_i,t})'$ is a $r_i$-dimensional latent factor process for the $i$-th group $\by_{i,t}$,  $\bA_i$ is the associated factor loading matrix, and $\be_{it}$ is the idiosyncratic term of the $i$-th group, which is independent of the group common factors $\bgg_{it}$.  Letting 
$\bgg_t:=(\bgg_{1,t}',...,\bgg_{m,t}')'$ be the collection of all group common factors, which is a $k_m$-dimensional latent vector with $k_m=r_1+...+r_m$, we assume that
\begin{equation}\label{global-f}
\bgg_t=\bB\bff_t+\bu_t,
\end{equation} 
where $\bff_t=(f_{1,t},...,f_{r,t})'$ is an $r$-dimensional global factor process capturing the cross-sectional dependence across the $m$ groups via their group factors, $\bB\in \mathbb{R}^{k_m\times r}$ is the 
associated loading matrix, and $\bu_t$ = $(\bu_{1t}',\ldots,\bu_{mt}')'$ is the idiosyncratic term of the stacked process $\bgg_t$ 
with $\bu_{it}$ constituting the {\em group specific factors} of the $i$-th group, which is orthogonal to the global factors $\bff_t$. Putting Equations (\ref{m-factor}) and (\ref{global-f}) together, 
we obtain the following matrix form of the proposed 2-level model: 
\begin{equation}\label{full:m}
\by_t=\bA\bB\bff_t+\bA\bu_t+\be_t,
\end{equation}
where $\bA=\diag(\bA_1,...,\bA_m)$ is a $p\times k_m$ block-diagonal matrix, $\bff_t$ is the global common factor process, and $\bu_t$ consists of $m$ vectors of group specific factor processes.  For the identification issue, we assume $\bff_t$, $\bu_t$, and $\be_t$ are mutually independent of each other. The decomposition in (\ref{global-f}) indicates that the global common factors come from the 
 cross-sectional dependence between the group factors, and the $\bu_t$ process, which captures less cross-sectional dependence, would be the specific factors uniquely representing the individual groups. 

We mention that \cite{andobai2017} and \cite{alonso2020} studied panel data with global and group-specific factors. However, our factors and the associated loading matrices are different from those of the aforementioned two papers. Most importantly, the methods used in the prior papers 
require the entire data set to estimate the global and group-specific factors, even when the group membership is known.

It is well known that the factors and their associated loading matrices in Models (\ref{m-factor}) and (\ref{global-f}) are not uniquely identified. We follow the restrictions used in \cite{BaiNg_Econometrica_2002} and \cite{fan2013} and assume that 
\begin{equation}\label{constr}
\Cov(\bgg_{i,t})=\bI_{r_i},\,\,\Cov(\bff_t)=\bI_r,\,\, \bA_i'\bA_i\,\,\text{and}\,\, \bB'\bB\,\,\text{are diagonal},\,\, i=1,...,m.
\end{equation}
Our goal here is to estimate $\bA_i$ and $\bB$, and to extract the factor 
processes $\bgg_{it}$ and $\bff_t$. We also study asymptotic properties of the proposed estimates.

\subsection{Estimation of Distributed Principal Components}
In this section, we discuss an estimation method based on 
a given realization $\{\by_t,|t=1,...,T\}$. 
We start with the case of knowing $r_i$ and $r$, where 
$r_i$ is the number of common factors of the $i$-th group and 
$r$ is the number of global common factors. 
The specifications of $r_i$ and $r$ are discussed shortly after. 
For each group (or machine), we use the least squares (LS) 
method to estimate $\bA_i$ and $\bgg_{i,t}$ as follows:
\begin{equation}\label{est-g}
(\wh\bA_i,\wh\bgg_{i,t})=\arg\min_{\bA_i\in \mathbb{R}^{p_i\times r_i}, \bgg_{i,t}\in \mathbb{R}^{r_i}}\sum_{t=1}^T\|\by_{i,t}-\bA_i\bgg_{i,t}\|_2^2,
\end{equation}
subject to the normalization conditions:
\begin{equation}\label{norm-g}
\frac{1}{T}\sum_{t=1}^T\bgg_{i,t}\bgg_{i,t}'=\bI_{r_i},\,\,\text{and}\,\, \bA_i'\bA_i\,\,\text{is diagonal}.
\end{equation}
Similarly to \cite{BaiNg_Econometrica_2002} and \cite{Bai_Econometrica_2003}, let $\bY_i=(\by_{i,1},...,\by_{i,T})'$ and $\bG_i=(\bgg_{i,t},...,\bgg_{i,T})'$. The Principal Components (PC) estimation is equivalent to the LS method in (\ref{est-g}) that minimizes $\tr[(\bY_i-\bG_i\bA_i')'(\bY_i-\bG_i\bA_i')]$ subject to the normalizations $\bG_i'\bG_i/T=\bI_{r_i}$ and $\bA_i'\bA_i$ being diagonal. Consider the eigenvalue-eigenvector analysis of the $T\times T$ matrix $\bY_i\bY_i'$. 
The estimator for $\bG_i$, denoted by $\wh\bG_i=(\wh\bgg_{i,1},...,\wh\bgg_{i,T})'$, is $\sqrt{T}$ times the eigenvectors corresponding to the $r_i$ largest eigenvalues, in decreasing order. Then, 
$\wh\bA_i=\bY_i'\wh\bG_i/T$ is the corresponding factor loading matrix. The common component matrix $\bG_i\bA_i'$ is estimated by $\wh\bG_i\wh\bA_i'$.

Turn to the estimations of $\bff_t$ and $\bB$. Let $\wh\bG=(\wh\bG_1,...,\wh\bG_m)$ and $\bF=(\bff_1,...,\bff_T)'$. We apply, again, the PC estimation method so that
\begin{equation}\label{est-c}
(\wh\bB,\wh\bF)=\arg\min_{\bB\in \mathbb{R}^{k\times r},\bF\in \mathbb{R}^{T\times r}}\|\wh\bG-\bF\bB'\|_F^2,
\end{equation}
subject to the constraints $\bF'\bF/T=\bI_r$ and $\bB'\bB$ being diagonal.
Therefore, $\wh\bF$ consists of the eigenvectors (multiplied by $\sqrt{T}$) associated with the $r$ largest eigenvalues, in decreasing order, of the $T\times T$ matrix $\wh\bG\wh\bG'$. Then, 
$\wh\bB=\wh\bG'\wh\bF/T$ and, hence, the group specific factors $\wh\bU=(\wh\bu_1,...,\wh\bu_T)'=\wh\bG-\wh\bF\wh\bB'$,  
where $\bu_t=(\bu_{1,t}',...,\bu_{m,t}')'$ with $\bu_{i,t}$ being the group-specific factor process associated with the $i$-th group.

Next, we discuss the determination of the number of factors in each group and the global one. For the estimation of $r_i$, we apply the well-known method in \cite{BaiNg_Econometrica_2002}. It estimates $r_i$ by
\begin{equation}\label{ri}
\wh r_i=\arg\min_{0\leq K_i\leq K}\log\left(\frac{1}{p_iT}\|\bY_i-\wh\bG_{i,K_i}\wh\bG_{i,K_i}'\bY_i/T\|_F^2\right)+K_ig(T,p_i),
\end{equation}
where $K$ is a prescribed upper bound, $\wh\bG_{i,K_i}$ is a $T\times K_{i}$ matrix whose rows are $\sqrt{T}$ times the eigenvectors corresponding to the $K_i$ largest eigenvalues of the $T\times T$ matrix $\bY_i\bY_i'$, and $g(T,p_i)$ is a penalty function of $(p_i,T)$ such that $g(T,p_i)=o(1)$ and $\min\{p_i,T\}g(T,p)\rightarrow\infty$. Two examples of $g(T,p_i)$ suggested by \cite{BaiNg_Econometrica_2002} are IC1 and IC2, respectively, 
\[g(T,p_i)=\frac{p_i+T}{p_iT}\log\left(\frac{p_iT}{p_i+T}\right)\quad \text{and}\quad g(T,p_i)=\frac{p_i+T}{p_iT}\log(\min\{p_i,T\}).\]
For the estimation of $r$, besides the information criterion as that in (\ref{ri}), we also adopt the ratio-based method in \cite{lamyao2012} and \cite{ahn2013} with some modifications. Let $\wh\lambda_1\geq ...\geq \wh\lambda_{k_m}$ be the $k_m$ eigenvalues of $\wh\bG'\wh\bG$, we estimate $r$ by
\begin{equation}\label{est-r}
\wh r=\arg\min_{1\leq l\leq R}{\frac{\wh\lambda_{l+1}}{\wh\lambda_{l}}},
\end{equation}
where we take $R=[k_m/2]$ as suggested by \cite{lamyao2012}. Alternatively, we may estimate the number of global factors by selecting a threshold $0<c_0<1$ such that
\begin{equation}\label{var-exp}
\wh r= \arg \min_j\left\{\frac{\sum_{l=1}^j\wh\lambda_{l}}{\sum_{l=1}^{k_m}\wh\lambda_{l}}\geq c_0\right\},
\end{equation}
where the ratio denotes the fraction of the variances explained by the first $j$ global factors. This 
estimate is easy to implement and could be practically useful for a prescribed $c_0$ (e.g., $c_0=85\%$).

\subsection{Clustering Time Series If Needed}
In this section, we consider the case in which the data need to be distributed to different machines by the central server according to certain characteristics of the observed time series, or the data are in small scale and stored on a single machine, but  require clustering. 
In other words,  we consider the case 
that the group memberships are not available. To apply the proposed method, we need to infer
 the group memberships of the series under study, which can be done via a clustering analysis.

Let $\by_t=(y_{1t},y_{2t},...,y_{pt})'$ be a $p$-dimensional time series of interest. 
There are many time-series clustering methods available in the literature, especially when 
the data can be stored in a single machine, and we briefly mention some of the commonly used ones. The first method is based on some selected features. Specifically, let $\bS=\{\bs_1,...,\bs_p\}$, where $\bs_i$ is a vector of 
selected features of the time series $\{y_{it},t=1,...,T\}$. 
For example, $\bs_i$ may contain the first few 
lags of the sample autocorrealtions or the sample partial autocorrelations  of $y_{it}$. Clustering can then be based on a chosen distance measure between $\bs_i$ and $\bs_j$. See, 
for instance, \cite{Caiado2006} and the references therein. The second method is based on a distance measure between the two series $y_{it}$ and $y_{jt}$ directly. For time series $\{y_{it}\}_{t=1}^T$ and $\{y_{jt}\}_{t=1}^T$, one can choose a distance measure $d_{ij}$, which can be the Euclidean or the Manhattan distance, to define the similarities, and the smaller the distance the closer the 
two series. The $k$-means clustering method can then be used. 
$d_{ij}$ can also be the cross-sectional correlations between the two series, and a large correlation implies that the two series can be grouped together. See, for 
instance, \cite{alonso2019} and the references therein.

When the dimension $p$ is sufficiently large so that the data cannot be stored in a single node, the above two classes of clustering methods have advantages and disadvantages. For clustering methods based on selected features: (a) There is no need to store all data in a single machine, but one needs to assume that the selected feature vectors $\bS$ can be stored in a single machine; (b) One may also need to transfer data between many computing units based on the clustering results obtained by the central server, which incurs communication costs. For clustering methods based on similarity measures: (a) One needs to compute the distance between all pairs of time series, which also requires  data transfer in and out of a single computing unit; (b) One also needs to transfer data to other machines based on the clustering results obtained by the central server. The communication costs can be 
extremely high due to the computation of the distances between all possible pairs of the time series.


In view of the above discussion, we leverage the advantages of feature-based procedures to 
propose a variation clustering procedure, which not only 
can distribute the series to different machines efficiently, but also avoids requiring to 
transfer the data to and out of the center server. 
The procedure is also motivated by the fact that PCA  is not scale invariant and often  allocates heavier weights  to series with larger sample 
variances. For example, we set the seed to \texttt{1234} in \texttt{R}  and simulated a 5-dimensional time series of a VAR(1) model with $500$ observations, where the coefficient 
matrix $\bPhi$ is generated by drawing elements of a $5\times 5$ matrix $\bM$ randomly from $U(-2,2)$, then properly normalized to ensure 
stationarity, and the noises are  independently generated from the standard multivariate normal distribution. Specifically, the VAR(1) coefficient matrix used is $\bPhi = 0.9\bM/\|\bM\|_2$. 
The variances of the five series are $(0.1484,0.1482,0.1574,0.1366,0.1582)$ so that they have similar variances. 
We apply PCA to these five time series and obtain the principal vectors below
\begin{equation*}\label{pca:v}
\bP_5=\left(\begin{array}{rrrrr}
0.451 &-0.332&  0.613  &0.093 & 0.550\\
-0.456 & 0.310 & 0.476& -0.670 & 0.144\\
-0.533 & 0.001 &  0.483 &  0.661 & -0.213\\
-0.147 & -0.833 &  0.084&  -0.316&  -0.423\\
 -0.532&  -0.318&  -0.396 &  0.078 &  0.673
\end{array}\right),
\end{equation*}
where the $i$-th column corresponds to the $i$-th principal direction in decreasing order of  the captured variances.
We now multiply the first series by 10 so that its variance becomes $14.84$. We apply PCA, again, 
and obtain the following principal vectors: 
\begin{equation*}\label{pca:vc}
\bP_5^*=\left(\begin{array}{rrrrr}
  0.999& -0.043 &  0.011  & 0.010&  -0.002\\
 -0.024 & -0.468 &  0.521&  -0.311&  -0.642\\
-0.025 &-0.604  &0.243 &-0.195 & 0.733\\
 0.001& -0.299 &-0.767& -0.550 &-0.139\\
-0.029& -0.570& -0.284 & 0.750 &-0.177\\
\end{array}\right).
\end{equation*}
Here the dynamical dependence between the five time series remains the same,  
but PCA assigns rather different weights to the first series in estimating the first principal component, as shown by the first column in $\bP_5$ and $\bP_5^*$.  This phenomenon 
may lead to complications in selecting and interpreting the resulting PCs and, hence, the common 
factors. To mitigate the scaling effect in applying PCA, 
let $\wt\by_t$ be the original $p$-dimensional series with unknown group memberships and $s_i^2$ be the sample variance of the $i$-th component, we require that $s_i^2$ be transferred to the 
central server, if needed. The central server then sorts the sample variances 
in the decreasing order as $s_{(1)}^2\geq s_{(2)}^2\geq...\geq s_{(p)}^2$. 
Denote the re-arranged series by $\by_t$ and we divide the $p$ series into $m$ groups 
of roughly equal sizes. That is, we  set $p_i= [ p/m]$ for $i=1,...,m-1$, 
and $p_m=p-(m-1)[ p/m]$, where $[p/m]$ is the integer part of $p/m$. 
The first $p_1$ series of $\by_t$ are assigned to the first node, the next $p_2$ series 
to the 2nd node and so on until the last $p_m$ series are assigned to the $m$-th node. 
In this way, the observed time series are assigned to groups based on their sample variances. 
 There are several advantages of this variation-based clustering method. 
 First, it can mitigate the scaling effect affecting the performance of PCA. 
 Second, it does not require transferring the entire time series data to the center server 
 and it is easy to implement, even when both the dimension $p$ and sample size $T$ are 
 reasonably large.

There are other efficient clustering methods available in the literature to assign individual 
time series to the $m$ nodes.  For instance, some prior (or domain) knowledge may be 
available for clustering.  The stock returns can naturally be clustered using their Standard Industrial Classification Codes, macroeconomic time series using their industrial characteristics, and spatio-temporal time series using their geographical locations. 
An important issue to consider is the cost of computing and data transfer  in the assignment.

 \subsection{Prediction}
 We introduce next a new variant of the Diffusion Index (DI) model to predict a selected series $x_t$ based on all the components of $\by_t$. In some applications, $x_t$ may be 
 associated with a specific group of $\by_t$. For example, in macroeconomic forecasting, $x_t$ can be classified into a  group of $\by_t$ according to its industrial sector; in spatio-temporal data, $x_t$ often belongs to a  cluster of $\by_t$ according to its geographic features. 
 Assume that $x_t$ belongs to the $i$-th 
 group. Then, with the estimated global factors $\wh\bff_t$ and the group-specific ones $\wh\bu_{i,t}$, the $h$-step ahead prediction  of $x_t$ at the forecast origin $t$ is
 \begin{equation}
\label{df-id}
\wh x_{t+h}=\wh\alpha_h+\wh\bbeta_h(L)\wh\bff_t+\wh\bgamma_h(L)\wh\bu_{i,t}+\wh\phi_h(L)x_{t}, 
 \end{equation}
 where 
 $\wh\bbeta_h(L)=\wh\bbeta_0'+\wh\bbeta_1'L+...+\wh\bbeta_{q_1}'L^{q_1}$, $\wh\bgamma_h(L)=\wh\bgamma_0'+\wh\bgamma_1'L+...+\wh\bgamma_{q_2}'L^{q_2}$,  
 and $\wh\phi_h(L)=\wh\phi_0+\wh\phi_1L+...+\wh\phi_{q_3}L^{q_3}$ are matrix or scalar 
 polynomials of order $q_1, q_2$ and $q_3$, respectively, with $L$ being the lag operator, $\wh\alpha_h, \wh\bbeta_i,\wh\bgamma_j$, and $\wh \phi_k$ are parameter estimates based on the available data $\{x_1,...,x_t\}$, $\{\wh\bff_1,...,\wh\bff_t\}$ and $\{\wh\bu_{i,1},...,\wh\bu_{i,t}\}$ using, for example, the ordinary LS method. On the other hand, if we are not certain about which group-specific factor process to use in (\ref{df-id}), we may include all the group-specific factors and using group penalized method to estimate the coefficients such as the group Lasso in \cite{yuanlin2006}. Here the cross-validation method 
 for choosing the tuning parameters should be replaced by a rolling-window method for time series data due to its dynamic structure. We do not pursue further any detail here.

 
\section{Theoretical Properties}\label{sec3}
We begin with some assumptions for Models (\ref{m-factor}) to (\ref{full:m}), where only $\{\by_{it}|1\leq i\leq m,1\leq t\leq T\}$ are observable. Similar ones are also given in \cite{BaiNg_Econometrica_2002} and \cite{fan2013}. All proofs of the theorems are given in 
the Appendix.

\begin{assumption}\label{asm1}
All the eigenvalues of $p_i^{-1}\bA_i'\bA_i$ and $k_m^{-1}\bB'\bB$ are bounded away from both $0$ and $\infty$, as $p_i, k_m\rightarrow\infty$.
\end{assumption}
\begin{assumption}\label{asm2}
The process $\{\by_{it},\bgg_{it}, \be_{i,t},\bff_{t}\}$ is stationary with zero mean, and $\alpha$-mixing with the mixing coefficient  $\alpha_p(k)\leq \exp(-Ck^{\gamma})$ for some positive constants
$C$ and $\gamma>0$, where
\[\alpha_p(k)=\sup_{i}\sup_{A\in\mathcal{F}_{-\infty}^i,B\in \mathcal{F}_{i+k}^\infty}|P(A\cap B)-P(A)P(B)|,
\]
and $\mathcal{F}_i^j$ is the $\sigma$-field generated by $\{(\by_{it},\bgg_{it}, \be_{i,t},\bff_{t}):i\leq t\leq j\}$.
\end{assumption}
\begin{assumption}\label{asm3}
(i) There exist constants $c_1,c_2>0$ such that $\lambda_{\min}(\bSigma_{e_i})>c_1$, $\lambda_{\min}(\bSigma_{u})>c_1$, $\|\bSigma_{e_i}\|_1<c_2$, and $\|\bSigma_u\|_1<c_2$, where $\bSigma_{e_i}=\Cov(\be_{i,t})$ and $\bSigma_u=\Cov(\bu_t)$;\\
(ii) There exists a constant $C>0$ such that $\bgg_{i,t}$, $\bff_t$, $\be_{i,t}$, and $\bu_t$ are sub-exponentially distributed in the sense that 
\[P(|\bv'(\bx_t-E(\bx_t))|>x)\leq C\exp(-Cx),\,\,\text{for}\,\, x>0,\]
where $\|\bv\|_2=1$ and $\bx_t$ is any vector of $\bgg_{i,t}$, $\bff_t$, $\be_{i,t}$, and $\bu_t$ .
\end{assumption}
\begin{assumption}\label{asm4}
There exists an $M>0$ such that, for all $l, k\leq p_i$, $j,\iota\leq k_m$, 
$s, t\leq T$, $1\leq i\leq m$,\\
(i) $\|\ba_{i,l}\|_\infty<M$ and $\|\bb_{j}\|_\infty<M$, where $\ba_{i,l}$ and $\bb_j$ are, respectively, the $l$-th and the $j$-th rows of $\bA_i$ and $\bB$; \\
(ii) $E\{p_i^{-1/2}[\be_{i,s}'\be_{i,t}-E(\be_{i,s}'\be_{i,t})]\}^4<M$, $E\{k_m^{-1/2}[\bu_{s}'\bu_{t}-E(\bu_{s}'\bu_{t})]\}^4<M$;\\
(iii) $E\|p_i^{-1/2}\sum_{l=1}^{p_i}\ba_{i,l}e_{i,lt}\|_2^4<M$ and $E\|k_m^{-1/2}\sum_{j=1}^{k_m}\bb_{j}u_{jt}\|_2^4<M$;\\
(iv) $\sum_{l=1}^{p_i}|E(e_{i,lt}e_{i,kt})|<M$ and $\sum_{j=1}^{k_m}|E(u_{jt}u_{\iota t})|<M$;\\
(v) $E\|\frac{1}{\sqrt{p_iT}}\sum_{t=1}^T\sum_{l=1}^{p_i}\bgg_{i,t}[e_{i,ls}e_{i,lt}-E(e_{i,ls}e_{i,lt})]\|_2^2<M$ and $E\|\frac{1}{\sqrt{k_mT}}\sum_{t=1}^T\sum_{j=1}^{k_m}\bff_{i,t}[u_{js}u_{jt}-E(u_{js}u_{jt})]\|_2^2<M$.
\end{assumption}
Assumption \ref{asm1} implies that the factors used are pervasive as those in \cite{BaiNg_Econometrica_2002} and \cite{fan2013}. Assumption \ref{asm2} is standard for dependent random processes, and it implies the temporal dependence conditions in \cite{BaiNg_Econometrica_2002} and \cite{Bai_Econometrica_2003}. See \cite{gaoetal2017} for a theoretical justification of the mixing conditions for VAR models. Assumption \ref{asm3}(i) guarantees that the covariances of the idiosyncratic terms are well defined, which will be used to show the consistency of the ratio-based method in (\ref{est-r}), and (ii) implies that the factors and the idiosyncratic terms all have exponential tails, which is stronger than the moment conditions in \cite{BaiNg_Econometrica_2002}.  Assumption \ref{asm4} controls the cross-sectional dependence of the idiosyncratic terms. 
Under Assumptions \ref{asm1}--\ref{asm3} and \ref{asm4}(i)-(iii), all the assumptions used in Bai and Ng (2002) are satisfied. See also \cite{fan2013}. Assumptions \ref{asm4}(iv)-(v) are imposed to control the cross-sectional dependence of the idiosyncratic terms as that of Assumption E(2) and Assumption F(1) in \cite{Bai_Econometrica_2003}, 
and the proof of Theorem 1 in \cite{BaiNg_joe_2006} is valid under these assumptions.



\begin{theorem}\label{tm1}
Suppose Assumptions \ref{asm1} to \ref{asm4} hold and $r_i$ and $r$ are known. Let  $\wh\bgg_{i,t}$ and $\wh\bff_{t}$ be the principal components estimators of $\bgg_{i,t}$ and $\bff_t$. Then, as $p_i, T\rightarrow\infty$, there exist full-rank matrices $\bH_i\in \mathbb{R}^{r_i\times r_i}$ and $\bK\in \mathbb{R}^{r\times r}$ such that
\begin{equation}\label{rate:g}
\frac{1}{T}\sum_{t=1}^T\|\wh\bgg_{i,t}-\bH_i'\bgg_{i,t}\|_2^2=O_p\left(\frac{1}{p_i}+\frac{1}{T}\right),\quad i=1,...,m
\end{equation}
and
\begin{equation}\label{rate:f}
\frac{1}{T}\sum_{t=1}^T\|\wh\bff_{t}-\bK'\bff_{t}\|_2^2=O_p\left(\frac{1}{k_m}+\frac{1}{T}+\frac{m}{k_m\underline{p}}+\frac{m^2}{{T^2}}+\frac{m^2}{{\underline{p}^2}}\right),
\end{equation}
where $\underline{p}=\min\{p_1,...,p_m\}$.
Furthermore, under the same assumptions, we have
\begin{equation}\label{max:rate1}
\max_{1\leq t\leq T }\|\wh\bgg_{i,t}-\bH_i'\bgg_{i,t}\|_2=O_p\left(\frac{1}{\sqrt{T}}+\frac{T^{1/4}}{\sqrt{p_i}}\right),
\end{equation}
and
\begin{equation}\label{max:rate2}
\max_{1\leq t\leq T }\|\wh\bff_{t}-\bK'\bff_{t}\|_2=O_p\left\{\sqrt{\frac{m}{k_m}}\log(T)\left[\frac{1}{\sqrt{T}}+\frac{1}{\sqrt{\underline{p}}}+\frac{m}{T}+\frac{m}{\underline{p}}\right]+\sqrt{\frac{m}{k_m}}\frac{T^{1/4}}{\sqrt{\underline{p}}}+\frac{T^{1/4}}{\sqrt{k_m}}\right\}.
\end{equation}
\end{theorem}

Some remarks of Theorem~\ref{tm1} are in order.
\begin{remark}\label{rm1}
(i) Theorem 1 implies that the convergence rates of the second step factor modeling depend on the number of clusters $m$, which is comparable with the rate of $k_m$ if the number of common factors in each group is small and fixed;\\
(ii) For the consistency of the estimated factors  in the sense of (\ref{rate:g}) and (\ref{rate:f}) as $p_i,T\rightarrow\infty$,  we require that  $m\rightarrow\infty$ but $m=o(T,\underline{p})$, which is a reasonable setting as the number of clusters is not extremely large in practice;\\
(iii) From (\ref{max:rate1}) and (\ref{max:rate2}), the uniform convergence can be achieved so long as $T=o(\underline{p}^2, m^2)$ and $m=o(T,\underline{p})$ as $p_i,T\rightarrow\infty$ under the assumption that $m=O(k_m)$. The conditions for the uniform convergence in (\ref{max:rate1}) and (\ref{max:rate2}) are stronger than those of (\ref{rate:g}) and (\ref{rate:f}) as they involve the maximums over the length of the time series $T$. However, the uniform rates do not provide a consistency pattern as $T\rightarrow\infty$, but $p_i$ and $m$ are large and fixed. We consider the discrepancies between the estimated loading matrices and factors with the true ones in Theorem \ref{tm3} below, which shows a consistency pattern as $T\rightarrow\infty$; \\
(iv) By \cite{BaiNg_joe_2006}, under the condition that all the eigenvalues of $p_i^{-1}\bA_i'\bA_i$ and $k_m^{-1}\bB'\bB$ are distinct, we can show that $\bH_i\rightarrow_p \bI_{r_i}$ and $\bK\rightarrow_p \bI_r$ up to a change of signs in the diagonals. We do not pursue this issue here.
\end{remark}

Partition $\bB=(\bB_1',...,\bB_m')'$ and let $\bC_i=\bA_i\bB_i$. Denote the $l$-th loading of $\bA_i$ and $\bC_i$ by $\ba_{i,l}$ and $\bfc_{i,l}$, respectively. We have the following theorem on the consistency of the estimated factor loading matrices.
\begin{theorem}\label{tm2}
Suppose Assumptions \ref{asm1} to \ref{asm4} hold, and $r_i$ and $r$ are known.  Then, for $1\leq i\leq m$,
\begin{equation}\label{max-loading1}
\max_{1\leq l\leq p_i}\|\wh\ba_{i,l}-\bH_i'\ba_{i,l}\|_2=O_p\left(\sqrt{\frac{\log(p_i)}{T}}+\frac{1}{\sqrt{p_i}}\right),
\end{equation}
and
\begin{equation}\label{max-loading2}
\max_{1\leq l\leq p_i}\|\wh\bfc_{i,l}-\bK'\bfc_{i,l}\|_2=O_p\left(\sqrt{\frac{1}{k_m}+\frac{m}{k_m\underline{p}}+\frac{m^2}{{T^2}}+\frac{m^2}{{\underline{p}^2}}}+\sqrt{\frac{\log(p_i)}{T}}\right), 
\end{equation}
where $\underline{p} = \min\{p_1,\ldots,p_m\}$. Consequently, 
\begin{equation}\label{max:gt}
\max_{l\leq p_i,t\leq T}\|\wh\ba_{i,l}'\wh\bgg_{i,t}-\ba_{i,l}'\bgg_{i,t}\|_2=O_p\left(\log(T)\sqrt{\frac{\log(p_i)}{T}}+\frac{T^{1/4}}{\sqrt{p_i}}\right),
\end{equation}
and
\begin{equation}\label{max:ft}
\max_{l\leq p_i, t\leq T}\|\wh\bfc_{i,l}'\wh\bff_{i,t}-\bfc_{i,l}'\bff_{i,t}\|_2=O_p\left(\log(T)w_T+\sqrt{\frac{m}{k_m}}\frac{T^{1/4}}{\sqrt{\underline{p}}}+\frac{T^{1/4}}{\sqrt{k_m}}\right),
\end{equation}
where $w_{T}=\sqrt{\frac{m}{k_m\underline{p}}+\frac{m^2}{{T^2}}+\frac{m^2}{{\underline{p}^2}}}+\sqrt{\frac{\log(p_i)}{T}}$. 
\end{theorem}
\begin{remark}
(i) Theorem 2 implies that, the uniform convergences of the estimated factor loading matrices and the extracted factors depend on the number of groups together with the dimension $p_i$ and the sample size $T$. Under the conditions specified in Remark \ref{rm1}, we can also achieve the consistency in Theorem \ref{tm2};\\
(ii) The convergence rate of $\|\wh\ba_{i,l}'\wh\bu_{i,t}-\ba_{i,l}'\bu_{i,t}\|_2$ follows from (\ref{max:gt}) and (\ref{max:ft}), and we do not repeat it here.
\end{remark}

Note that $\bA_i$ and $\bB$ cannot be uniquely determined due to the identification issue. But the linear spaces spanned by their columns, denoted by $\mathcal{M}(\bA_i)$ and $\mathcal{M}(\bB)$,  respectively, can be uniquely estimated. See \cite{gaotsay2020a,gaotsay2020b} for details. 
To this end, we adopt the discrepancy measure used by
\cite{panyao2008}, which is a well-defined distance measure on a quotient matrix space as shown therein: for two $p\times d$ semi-orthogonal
matrices ${\bf Q}_1$ and ${\bf Q}_2$ satisfying the condition ${\bf
Q}_1'{\bf Q}_1={\bf Q}_2'{\bf Q}_2=\bI_{d}$, the difference
between the two linear spaces $\mathcal{M}({\bf Q}_1)$ and
$\mathcal{M}({\bf Q}_2)$ is measured by
\begin{equation}
D({\bf Q}_1,{\bf
Q}_2)=\sqrt{1-\frac{1}{d}\textrm{tr}({\bf Q}_1{\bf Q}_1'{\bf
Q}_2{\bf Q}_2')}.\label{eq:D}
\end{equation}
Note that $D({\bf Q}_1,{\bf Q}_2) \in [0,1].$ 
It is equal to $0$ if and only if
$\mathcal{M}({\bf Q}_1)=\mathcal{M}({\bf Q}_2)$, and to $1$ if and
only if $\mathcal{M}({\bf Q}_1)\perp \mathcal{M}({\bf Q}_2)$.  In addition, if we denote the singular values of $\bQ_1'\bQ_2$ by $\{\sigma_i\}_{i=1}^d$ in descending order, then the principal angles between $\mathcal{M}(\bQ_1)$ and $\mathcal{M}(\bQ_2)$, denoted by $\bTheta(\bQ_1,\bQ_2)=\diag(\theta_1,...,\theta_d)$, are defined as $\diag\{\cos^{-1}(\sigma_1),...,\cos^{-1}(\sigma_d)\}$; see, for example,  Theorem I.5.5 of \cite{stewart-sun1990}. The squared Frobenius norm of the so-called $\sin\bTheta$ distance is 
\begin{equation}\label{sin:t}
\|\sin\bTheta(\bQ_1,\bQ_2)\|_F^2:=\sum_{i=1}^d\sin^2(\theta_i)=\sum_{i=1}^d(1-\sigma_i^2)=d\{D({\bf Q}_1,{\bf
Q}_2)\}^2,
\end{equation}
which implies that the above two measures are equivalent for finite $d$.
Therefore, we shall use only the distance in (\ref{eq:D}) to present our theoretical results.
The following theorem establishes the consistency of the estimated loading matrix $\wh\bA_i$,  the matrix $\wh\bB$, and the extracted common factors $\wh\bA_i\wh\bgg_{i,t}$ and $\wh\bA_i\wh\bB_i\wh\bff_t$.
 \begin{theorem}\label{tm3}
 Under the assumptions of Theorem \ref{tm1}, for $1\leq i\leq m$, we have
 \begin{equation}\label{DA:DAB}
 D(\wh\bA_i,\bA_i)=O_p(T^{-1/2})\quad \text{and}\quad D(\wh\bA\wh\bB,\bA\bB)=O_p\left(mT^{-1/2}+\underline{p}^{-1}\bar{p}T^{-1/2}+\underline{p}^{-1}\right).
 \end{equation}
 Furthermore, 
 \[p_i^{-1/2}\|\wh\bA_i\wh\bgg_{i,t}-\bA_i\bgg_{i,t}\|_2=O_p\left(p_i^{-1/2}+T^{-1/2}\right),\]
 and
 \[p^{-1/2}\|\wh\bA\wh\bB\wh\bff_t-\bA\bB\bff_t\|_2=O_p\left(p^{-1/2}k_m^{1/2}\delta_T+p^{-1/2}\right),\]
where $\bar{p}=\max\{p_1,...,p_m\}$, $\underline{p}=\min\{p_1,...,p_m\}$,  and $\delta_T=mT^{-1/2}+\underline{p}^{-1}\bar{p}T^{-1/2}+\underline{p}^{-1}$.
 \end{theorem}
 Theorem \ref{tm3} explores a different aspect of the consistency from that in Theorem \ref{tm2}. The uniform rates in Theorem \ref{tm2} do not imply the convergence pattern as $T\rightarrow\infty$ for fixed but large $p_i$, while the average consistencies in Theorem \ref{tm3} provide the asymptotic behavior of the estimated loading matrices and the extracted factors, and they are indeed consistent as the sample size $T\rightarrow\infty$, $\underline{p} \rightarrow \infty$,  $m=o(\sqrt{T})$, and the clusters are of similar sizes. The condition for $m$ is different from that in Theorem 1, but we can achieve the consistencies in (\ref{rate:g}) and (\ref{rate:f}) so long as  $m=o(\sqrt{T},\underline{p})$. The uniform consistency in  (\ref{max:rate2}) can still be achieved under this setting so long as $T^{1/4}/\sqrt{k_m}\rightarrow\infty$ if some $r_i$ diverges slowly. On the other hand, the conditions of $m$ in Theorem \ref{tm3} can be relaxed, because we can apply the Bernstein-type inequality in, for example, \cite{merlevede2011}, to reduce the term $m$ in (\ref{DA:DAB}) to an exponent of $\log(m)$   under sightly stronger conditions such that $\max_{1\leq i\leq m}D(\wh\bA_i,\bA_i)=O_p(\sqrt{\log(m)/T})$, instead of $O_p(mT^{-1/2})$. We omit the  details here to save space.
 

Next, we discuss the selections of the number of factors as they are unknown in practice. 
\begin{theorem}\label{tm4}
Under the assumptions of Theorem \ref{tm1}, $P(\wh r_i=r_i)\rightarrow 1$ and $P(\wh r=r)\rightarrow 1$, as $p_i, T\rightarrow\infty$, where $\wh r_i$ and $\wh r$ are defined in (\ref{ri}) and (\ref{est-r}), respectively.
\end{theorem}
By Theorem \ref{tm4}, we can replace $r_i$ and $r$ in Theorems \ref{tm1} and \ref{tm2} by $\wh r_i$ and $\wh r$, respectively, and the results therein continue to hold. See \cite{gaotsay2020b} for a similar argument. For the discrepancy measure defined in (\ref{eq:D}) and those considered in Theorem \ref{tm3}, we take $\wh\bA_i$ for example and  modify (\ref{eq:D}) as
\[\wt D(\wh\bA_i,\bA_i)=\sqrt{1-\frac{1}{\max(\wh r_i,r_i)}\textrm{tr}({\wh\bA}_i{\wh\bA}_i'{\bA}_i{\bA}_i')},\]
which takes into account the fact that the dimensions of $\wh\bA_i$ and $\bA_i$ may be different. See \cite{gaotsay2020b} for a similar argument.

 \section{Extension to Unit-Root Nonstationary Time Series}\label{sec4}
 
 In this section, we consider the case when the series $\by_{i,t}$ is an $I(1)$  process, for $1\leq i\leq m$, and, hence,  extend the proposed method to  unit-root nonstationary time series. To do so, we assume that $\bff_t$ in (\ref{global-f}) is an $I(1)$ process that captures the global stochastic trends, and $\be_{i,t}$ and $\bu_t$ are stationary components. Consequently, each $\by_{i,t}$ is an $I(1)$ process and $\bgg_{i,t}$ is the group common stochastic trend of the $i$-th group. This setup is natural and is similar to the nonstationary factor models of  \cite{penaponcela2006}, in which the dimension is fixed and $\bu_{i,t}$ is the stationary common factor of the $i$-th group. However,  we do not need the idiosyncratic terms $\be_{i,t}$ to be white noises. Under the assumption, the first PCA conducted in each group is to seek the group common stochastic trends, and the second one tries to find the global common stochastic trends and, hence, allows for the existence of co-integration relationships between the group common factors. On the other hand, as the dimension of  $\bgg_t$ is relatively large, 
 we consider the case  that its cointegration rank is low. Therefore, we assume that $r_i\leq r$ in the sequel, implying that the number of  global common trends can be relatively large. 
 
Although the estimation procedures used to obtain the common stochastic trends and their associated loading matrices  remain valid  for unit-root series in finite samples, except for the methods to determine the number of stochastic trends,  we  introduce another set of identification conditions to maintain the dynamic patterns of all unit-root common factors. Specifically, in the estimation, we assume
 \begin{equation}\label{non:cond}
 \bA_i'\bA_i=\bI_{r_i}\quad\text{and}\quad\bB'\bB=\bI_{r}.
 \end{equation}
 The above identification conditions and Assumption \ref{asm1} imply that an additional strength of order $p_i^{1/2}$ (and $k_m^{1/2}$) is imposed on $\bgg_{i,t}$ (and $\bff_{t}$). The estimation of $\bA_i$ is based on the sample covariance 
$$\wh\bSigma_{y_i}=\frac{1}{T}\sum_{t=1}^T(\by_{i,t}-\bar{\by}_i)(\by_{i,t}-\bar{\by}_i)',$$
where $\bar{\by}_i=T^{-1}\sum_{t=1}^T\by_{i,t}$. It is a scaled version of $\bY_i'\bY_i$ in Section 2 if the data are centered first.
 Under Assumption \ref{asm3}, by Theorems 1 and 2  of \cite{penaponcela2006}, we have 
 \begin{equation}\label{eigen}
 0\leq \ba_{i,j}'\wh\bSigma_{y_i}\ba_{i,j}\leq\left\{ \begin{array}{cc}
 CT&\text{if},\, E(\ba_{i,j}'\by_{i,t})=0,\\
 CT^2&\text{if}\,\, E(\ba_{i,j}'\by_{it})\neq 0,
 \end{array}\right.
 \end{equation}
 for any column vector $\ba_{i,j}$ of $\bA_i$ and  some constant $0<C<\infty$. The rate depends on whether there is a drift or not in the unit-root series. On the other hand, the constraints in (\ref{non:cond}) imply that 
the eigenvalues of $\wh\bSigma_{g_i}=T^{-1}\sum_{t=1}^T(\bgg_{it}-\bar{\bgg}_i)(\bgg_{it}-\bar{\bgg}_i)'$ is bounded below by $Cp_iT$ or $Cp_iT^2$ depending on whether $E(\bff_t)=
\mathbf{0}$ or not. We assume in the section that  $E(\bff_t)=\mathbf{0}$ for simplicity.

We first assume that $r_i$ and $r$ are known. Similarly to the argument in \cite{Harris1997} that the $I(1)$ series components of $\bgg_{i,t}$ are the linear combinations of $\by_{i,t}$ that yield explosive variances, and the ones with minimum sample variances would be the ``most stationary" components. Therefore, the estimated loading matrix $\wh\bA_i$ consists of the eigenvectors of $\wh\bSigma_{y_i}$ corresponding to the $r_i$ largest eigenvalues. The estimated factors are $\wh\bgg_{i,t}=\wh\bA_i'\by_{i,t}$. Although the way to obtain $\wh\bA$ and $\wh\bgg_{i,t}$ is different from the method of Section 2, but the extracted factor term $\wh\bA_i\wh\bgg_{i,t}$ remains the same due to the identification issue.

Next, define the normalized group common factors $\wh\bgg_{i,t}^*=\wh\bgg_{i,t}/\sqrt{p_i}$ and the stacked group factors $\wh\bgg_t^*=(\wh\bgg_{1,t}^{*'},...,\wh\bgg_{m,t}^{*'})'$. The estimation of $\bB$ is based on the following sample covariance matrix
\begin{equation}\label{cov:gs}
\wt\bSigma_{g}=\frac{1}{T}\sum_{t=1}^T(\wh\bgg_t^*-\bar{\wh\bgg}^*)(\wh\bgg_t^*-\bar{\wh\bgg}^*)',
\end{equation}
 where $\bar{\wh\bgg}^*=T^{-1}\sum_{t=1}^T\wh\bgg_t^*$. Consequently, the estimator $\wh\bB$ consists of the $r$ eigenvectors of $\wt\bSigma_{g}$ associated with the $r$ largest eigenvalues. The resulting estimated global stochastic trend is $\wh\bff_t=\wh\bB'\wh\bgg_t^*$.
 
 To estimate the number of common trends, 
 we adopt the auto-correlation based method in \cite{gaotsay2020a}. Specifically, for some prescribed integer $\bar{k}>0$, define
 \begin{equation}\label{auto:cor}
 S_{i,l}(\bar{k})=\sum_{k=1}^{\bar{k}}|\wh\rho_{i,l}(k)|,\,\, \text{and}\,\,S_j(\bar{k})=\sum_{k=1}^{\bar{k}}|\wh\rho_j(k)|,
 \end{equation}
 where $\wh\rho_{i,l}(k)$ is the lag-$k$ sample autocorrelation function (ACF) of the principal components $\wh g_{i,lt}$, for $1\leq i\leq m$ and $1\leq l\leq p_i$, and $\wh\rho_j(k)$ is the corresponding one of $\wh f_{j,t}$, for $1\leq j\leq k_m$. If $\wh g_{i,lt}$ (or $\wh f_{j,t}$) is stationary, then under some mild conditions, $\wh\rho_{i,l}(k)$ (or $\wh\rho_j(k)$) decays to zero  exponentially as $k$ increases, and $\lim_{\bar{k}\rightarrow\infty} S_{i,l}(\bar{k})<\infty$  (or $\lim_{\bar{k}\rightarrow\infty} S_{j}(\bar{k})<\infty$)  as $T\rightarrow\infty$.
 If $\wh g_{i,lt}$ (or $\wh f_{j,t}$) is unit-root nonstationary, then $\wh\rho_{i,l}(k)\rightarrow 1$ (or $\wh\rho_j(k)\rightarrow 1$), and $\lim_{\bar{k}\rightarrow\infty} S_{i,l}(\bar{k})=\infty$  (or $\lim_{\bar{k}\rightarrow\infty} S_{j}(\bar{k})=\infty$)  as $T\rightarrow\infty$. Therefore, for the $i$-th group, we start with $l=1$. If the average of the absolute sample ACFs $S_{i,l}(\bar{k})/\bar{k}\geq \delta_0$ for some constant $0< \delta_0<1$, then $\wh g_{i,lt}$ has a unit root and we increase $l$ by $1$ to repeat the detecting process. This detecting process is continued until $S_{i,l}(\bar{k})/\bar{k}< \delta_0$ or $l=p_i$. If $S_{i,l}(\bar{k})/\bar{k}\geq \delta_0$ for all $l$, then $\wh r_i=p_i$; otherwise, we denote $\wh r_i=l-1$. The estimate $\wh r$ of the 
 number of global common trends can similarly be obtained.

To study the theoretical properties of the estimators of the unit-root processes, we make some additional assumptions. Let $\bw_t=(w_{1t},...,w_{rt})'=\bff_t-\bff_{t-1}$ with $E(\bw_t)=\bf 0$, and define $\bS_T^r(\bt)=(S_T^1(t_1),...,S_T^r(t_r))'=\left(\frac{1}{\sqrt{n}}\sum_{s=1}^{[Tt_1]}w_{1s},...,\frac{1}{\sqrt{T}}\sum_{s=1}^{[Tt_{r}]}w_{rs}\right)'$,  where $0< t_1<...<t_r\leq 1$ are constants and $\bt=(t_1,...,t_{r})'$. 
\begin{assumption}\label{asm5}
(i) The process $\{\bw_t,\be_{i,t},\bu_t\}$ is $\alpha$-mixing with the mixing coefficient satisfying the condition $\alpha_p(k)\leq \exp(-ck^{\gamma})$ for some constants $c>0$ and $\gamma$ given in Assumption \ref{asm2}; \\
(ii) $\bw_t,\be_{i,t}$ and $\bu_t$ are sub-exponentially distributed as given in Assumption 3(ii);\\
(iii) There exists a Gaussian process $\bW(\bt)=(W_1(t_1),...,W_{r}(t_{r}))'$ such that as $T\rightarrow \infty$,
$\bS_T^{r}(\bt)\overset{J_1}{\Longrightarrow}\bW(\bt)\,\, \text{on}\,\, D_{r_1}[0,1],$
where $\overset{J_1}{\Longrightarrow}$ denotes weak convergence under the Skorokhod $J_1$ topology, and $\bW({\bf 1})$ has a positive definite covariance matrix $\bOmega=[\sigma_{ij}]$.
\end{assumption}
 \begin{assumption}\label{asm6}
 There exists a constat $\delta$ with $0<\delta<1/2$ such that
 \[\frac{1}{T^{1+\delta}}\sum_{t=1}^T f_{j,t} u_{k,t}=o_p(1)\quad \text{and}\quad \frac{1}{T^{1+\delta}}\sum_{t=1}^Tf_{j,t}e_{i,\iota t}=o_p(1),\]
 uniformly, for $1\leq j\leq r$, $1\leq k\leq k_m$,  $1\leq i\leq m$, and\, $1\leq \iota\leq p_i$.
 \end{assumption}
 Assumptions \ref{asm5}(i)-(ii) are used to replace the ones in Assumptions \ref{asm2} and \ref{asm3}(ii) and Assumption \ref{asm5}(iii) is standard for unit-root processes. Assumption \ref{asm6} is not strong and it can be established under the setting of \cite{Stock1987}, where we can assume  $\by_{i,t}$ and $\bgg_t$ have similar structures as the one in (2.4) of \cite{Stock1987} and there are cointegration relationships among them. 
 
 We now  state the convergence of the linear spaces spanned by the columns of the estimated loading matrices. Similar results as those in Theorems \ref{tm1} and \ref{tm2} continue to hold with a modified normalization in the proofs; see also \cite{bai2004} for a similar argument regarding the theoretical results with stochastic trend factors and the stationary ones. We do not repeat them here.
 \begin{theorem}\label{tm5}
 Suppose Assumptions \ref{asm1}, \ref{asm3}(i), \ref{asm4}(i) to (iv), \ref{asm5}, and \ref{asm6}  hold, and $r_i$ and $r$ are known, Then, for $1\leq i\leq m$,
 we have
 \begin{equation}\label{DA:DAB-1}
 D(\wh\bA_i,\bA_i)=O_p\left(\frac{1}{T^{1-\delta}}\right)\quad \text{and}\quad D(\wh\bA\wh\bB,\bA\bB)=O_p\left(\frac{m}{T^{1-\delta}}+\frac{\bar{p}}{T^{1-\delta}\underline{p}}\right).
 \end{equation}
 Furthermore, 
 \[p_i^{-1/2}\|\wh\bA_i\wh\bgg_{i,t}-\bA_i\bgg_{i,t}\|_2=O_p\left(p_i^{-1/2}+T^{-1/2+\delta}\right),\]
 and
 \[p^{-1/2}\|\wh\bA\wh\bB\wh\bff_t-\bA\bB\bff_t\|_2=O_p\left(p^{-1/2}k_m^{1/2}\eta_T+p^{-1/2}\right),\]
where $\bar{p}=\max\{p_1,...,p_m\}$, $\underline{p}=\min\{p_1,...,p_m\}$, and $\eta_T=mT^{-1/2+\delta}+\underline{p}^{-1}\bar{p}T^{-1/2+\delta}$.
 \end{theorem}
 \begin{remark}
 (i) The convergence rate of $\wh\bA_i$ is informative. In fact, we can drop the small constant $\delta$ if we make an assumption similar to the distribution results in \cite{Stock1987} in which the sample statistics in Assumption \ref{asm6} are stochastically bounded with $\delta=0$. The resulting rate is $T$, which is in agreement with that of the estimated cointegrating vector in the literature;\\
 (ii) The second rate in (\ref{DA:DAB-1}) depends on the number of nodes $m$ and the ratio between the largest and the smallest group dimensions. The constant $m$ can be reduced to a smaller one as we applied the Bonferroni's inequality to the $m$ groups in the proof. If $m$ is reduced and the dimensions of the groups are comparable, then the convergence rate will be  similar to that of $\wh\bA_i$.
 \end{remark}

The consistency of the estimated numbers of the group stochastic trends and the global trends using the average of the absolute autocorrelations   can be shown in a similar way as that in  \cite{gaotsay2020a} and the details are omitted. 

\section{Incorporating the Horizontal Partition}\label{sec5}

In this section, we briefly discuss the extreme case  in which the sample size $T$ is too large for 
a single machine to store the entire data of the $p_i$ time series, leading to the need to 
partition the data of those $p_i$ series. This situation is referred to as 
the  horizontal-partition regime in the {\em Introduction}. Again, for simplicity, we 
 focus on the 2-level case in Figure~\ref{fig0}, but assume that additional $n$ new machines are affiliated with each of the $m$ nodes, where $n$ depends implicitly on the sample size $T$, 
the number of nodes $m$ and the dimension $p_i$. We assume that the sample size $T$ is 
partitioned, in time order,  as $T=T_1+\cdots+T_n$ such that the $T_j$ observations of  the $p_i$-dimensional series can be analyzed by a single machine, for $i=1,\ldots,m$ and $j=1,\ldots,n$.  
Thus, we partition the $p_i$ dimensional series into $n$ blocks as 
 $\{\by_{i,jt}|t=1,...,T_j; j=1,\ldots, n\}$, for $i=1,\ldots, m$, and store 
the data $\{\by_{i,jt}|t=1,\ldots,T_j\}$ in the $j$-th new machine affiliated with the $i$-th node. 
Assuming further that all the numbers of common factors $r_i$ and $r$ are known, 
we introduce next two approaches to 
estimating the common factors and their associated loading matrices.

The first approach applies the proposed method of Section \ref{sec2} to each sub-sample $\{\by_{i,jt},t=1,...,T_j\}$ and obtains the estimated loading matrix and factors, say, $\wh\bA_{i,j}$ and $\{\wh\bgg_{i,jt},t=1,...,T_j\}$, for $j=1,\ldots, n$ and $i=1,\ldots,m$. 
Then we collect all the factors resulted from the $j$-th machine associated with the $i$-th node and form $\wh\bG_{i,j}=(\wh\bgg_{i,j1},...,\wh\bgg_{i,jT_j})'$. Define $\wh\bG_j^n=(\wh\bG_{1,j},...,\wh\bG_{m,j})$, which consists of all the group factors for the $j$-th segment, where the 
subscript $n$ signifies the partition of $n$ blocks. Next, we apply PCA 
again to $\wh\bG_j^n$ as in (\ref{est-c}), and denote the estimated loading matrix and factors by 
$\wh\bB_j^n$ and $\{\wh\bff_{j,t}: t=1,...,T_j\}$, respectively, which correspond to the $j$-th segment. Consequently, the estimated group common factors and global factors are $\{\wh\bgg_{i,11,},...,\wh\bgg_{i,1T_1},...,\wh\bgg_{i,n1},...,\wh\bgg_{i,nT_n}|i=1,...,m\}$ and $\{\bff_{1,1},...,\bff_{1,T_1},...,\bff_{n,1},...,\bff_{nT_n}\}$, respectively.


Let $\wh\bA_j^n=\diag(\wh\bA_{1,j},...,\wh\bA_{m,j})$ be the block diagonal matrix consisting of all the estimated loading matrices from the $j$-th segment. Then, 
$\wh\bA_j^n$ is an estimate of $\bA$ and $\wh\bB_j^n$ is an estimate of $\bB$ based on the $j$-th segment. By Theorem \ref{tm3}, 
 we have the following corollary regarding the consistency of the loading matrices.
\begin{corollary}
If the conditions of Theorem \ref{tm1} hold for the $j$-th segment of the data. Then, for $1\leq i\leq m$, and  $1\leq j\leq n$, we have
 \begin{equation}\label{DA:DAB-2}
 D(\wh\bA_{i,j},\bA_i)=O_p(T_j^{-1/2})\quad \text{and}\quad D(\wh\bA_j^n\wh\bB_j^n,\bA\bB)=O_p(mT_j^{-1/2}+\underline{p}^{-1}\bar{p}T_j^{-1/2}+\underline{p}^{-1}).
 \end{equation}
\end{corollary}
In fact, all the theoretical properties in Theorems \ref{tm1} to \ref{tm3} remain valid for the $j$-th segment so long as we replace $T$ therein by $T_j$. A key consideration for using each data 
segment to estimate $\bA$ and $\bB$ is that, in real applications with extremely long time series,  the stationarity of $\by_{j,t}$ may fail. If this occurs,  our first approach can extract more accurately  the common factors using $\wh\bA_{i,j}$ for the $j$-th segment rather than using a globally estimated loading  matrix $\wh\bA_i$.

The second approach is similar to that in \cite{fan2019} of the horizontal-partition regime.
We adopt the identification conditions that $\bA_i'\bA_i=\bI_{r_i}$ and $\bB'\bB=\bI_r$. Letting $\bSigma_{y_i,j}=\Cov(\by_{i,jt})$ and $\wh\bSigma_{y_i,j}$ be its sample estimate, we denote $\wh\bA_{i,j}$ the matrix consisting of the  eigenvectors of $\wh\bSigma_{y_i,j}$ associated with the $r_i$ largest eigenvalues. The estimated loading $\wh\bA_i$ for the entire data 
set of the $p_i$-dimensional time series is the matrix consisting of the eigenvectors 
associated with the  $r_i$ largest eigenvalues of 
\[\wt\bSigma_{y_i}^n=\frac{1}{n}\sum_{j=1}^n\wh\bA_{i,j}\wh\bA_{i,j}',\]
which is similar to the model averaging technique in machine learning, and 
we compute the factors $\{\wh\bgg_{i,t},t=1,...,T\}$ by making use of the final estimator $\wh\bA_i$ for the $i$-th group. 

Next, we partition $\{\wh\bgg_{i,t},t=1,...,T\}$ into $n$ segments and stack all the $j$-th segment associated with the $i$-th group to obtain $\wh\bG_j^n$ as before. Similarly to the way we obtain $\wh\bA_i$, letting $\wh\bB_j^n$ be the matrix consisting of the eigenvectors of the $r$ largest eigenvalues 
of $\wh\bG_j^n{'}\wh\bG_j^n/T_j$, we perform eigenvalue-eigenvector analysis on 
\[\wt\bSigma_{g}^{n}=\frac{1}{n}\sum_{j=1}^n\wh\bB_j^n\wh\bB_j^n{'},\]
and the final estimated loading matrix $\wh\bB$ consists of the eigenvectors associated 
with the $r$ largest eigenvalues of $\wt\bSigma_{g}^{n}$.

Although the second procedure is similar to that in \cite{fan2019}, the theoretical properties 
are much more involved as the data are time series with dynamically dependent structures and the theoretical results in \cite{fan2019} are only valid for i.i.d. data. Furthermore, the data involved in the second-step PCA may not be weakly dependent as $\wh\bA_i$ contains information across the entire time span. Therefore, the bias-variance argument in \cite{fan2019} is no longer valid under the 
current framework. We leave details of this approach for future research.

\section{Numerical Results}\label{sec6}
We use simulation and a real example to assess the performance of the proposed analysis in finite samples.

\subsection{Simulation}
We study the finite-sample properties of the proposed methodology under the scenarios when $p$ is both small and large. 
As the dimensions of $\wh\bA_i$  and $\bA_i$ are not necessarily the same,
and  $\bA_i$ is not necessarily an orthogonal matrix, we extend the discrepancy measure
in Equation (\ref{eq:D}) to a more general one given below. Let $\bQ_i$ be a
$p\times d_i$ matrix with rank$(\bQ_i) = d_i$, and $\bP_i =
\bQ_i(\bQ_i'\bQ_i)^{-1} \bQ_i'$, $i=1,2$, where we use the Moore-Penrose generalized inverse. Define
\begin{equation}\label{dmeasure}
\bar{D}(\bQ_1,\bQ_2)=\sqrt{1-
\frac{1}{\max{(d_1,d_2)}}\textrm{tr}(\bP_1\bP_2)},
\end{equation}
then $\bar{D} \in [0,1]$. Furthermore,
$\bar{D}(\bQ_1,\bQ_2)=0$ if and only if
either $\mathcal{M}(\bQ_1)\subset \mathcal{M}(\bQ_2)$ or
$\mathcal{M}(\bQ_2)\subset \mathcal{M}(\bQ_1)$, and  it is 1 if and only if
$\mathcal{M}(\bQ_1) \perp \mathcal{M}(\bQ_2)$.
When $d_1 = d_2=d$ and $\bQ_i'\bQ_i= \bI_d$,
$\bar{D}(\bQ_1,\bQ_2)$ reduces to $D(\bQ_1,\bQ_2)$ of Equation (\ref{eq:D}). 

In the simulation study, we mainly focus on the numerical performance of the proposed method in the vertical partition regime for stationary and unit-root non-stationary time series data, because, in a  horizontal partition regime, each block will produce  similar results as those shown below, and we omit the details here to save space.

\noindent {\bf Example 1.}  Consider the models in (\ref{m-factor}) and (\ref{global-f}) with 
the global common factors following the VAR(1) model
\[\bff_t=\bPhi\bff_{t-1}+\bfeta_t,\]
where $\bfeta_t$ is a white noise process and $\bPhi$ is a diagonal matrix. We set the number of groups $m=10,\ 20$. For each $m$, the true number of global factors is $r=3$, the numbers of group factors are $r_1=...=r_m=5$, the group dimensions are $p_1=...=p_m=20,30,40$, or 50, and the sample size is $T=400,\ 800$, or 1200. For the coefficient matrices, we first set the seed in \texttt{R} to \texttt{1234} and the elements of $\bB$ and $\bA_{i}$ are drawn independently from $U(-2,2)$, and the diagonal elements of $\bPhi$ are drawn independently from $U(0.5,0.9)$. 
For each realization of $\by_t$,  $\bfeta_t \sim N(0,\bI_{r})$, $\bu_t\sim N(0,\bI_{k_m})$, and $\be_{i,t}\sim N(0,\bI_{p_i})$. We use $500$ replications for each $(p_i,m,T)$ configuration, where all the $p_i$'s are set to be equal.

We first consider the performance of estimating the number of factors, and  use the IC1 criterion with $K=[p_i/3]$ to determine $r_i$ and the ratio-based method in (\ref{est-r}) with $R=[k_m/2]$ to estimate $r$ . The empirical probabilities of $\min_{1\leq i\leq m}P(\wh r_i=r_i)$ and $P(\wh r=r)$ are reported in Table \ref{Table1} for $m=10$ and $20$. From the table, we see that 
the proposed methods can successfully estimate the numbers of group factors using the BIC and 
that of the global factors using the ratio-based procedure. This  is understandable because the factors 
used are all strong ones. Similar results can also be found in \cite{BaiNg_Econometrica_2002} and \cite{lamyao2012}.

\begin{table}
 \caption{Empirical probabilities $\min_{1\leq i\leq m}EP(\hat{r}_i=r_i)$, 
 denoted by $\min_{1\leq i\leq m}EP_i$, and $EP(\wh r=r)$ of various $(p_i,m,T)$ configurations 
 for the model of Example 1 with $r_i=5$ and $r=3$, where $p_i$ and $T$ are, respectively,  the dimension and the sample size of each group. $500$ iterations are used.} 
          \label{Table1}
\begin{center}
 \setlength{\abovecaptionskip}{0pt}
\setlength{\belowcaptionskip}{3pt}

\begin{tabular}{c|ccccccccc}
\hline
\hline
&&\multicolumn{2}{c}{$T=400$}&&\multicolumn{2}{c}{$T=800$}&&\multicolumn{2}{c}{$T=1200$}\\
$m=10$&&${\min\atop{1\leq i\leq m}}EP_i$&$EP(\wh r=r)$&&${\min\atop{1\leq i\leq m}}EP_i$&$EP(\wh r=r)$&&${\min\atop{1\leq i\leq m}}EP_i$&$EP(\wh r=r)$\\
\cline{1-2}\cline{3-4}\cline{6-7}\cline{9-10}
$p_i=20$&&1&1&&1&1&&1&1\\
$p_i=30$&&1&1&&1&1&&1&1\\
$p_i=40$&&1&1&&1&1&&1&1\\
$p_i=50$&&1&1&&1&1&&1&1\\
\hline
\hline
&&\multicolumn{2}{c}{$T=400$}&&\multicolumn{2}{c}{$T=800$}&&\multicolumn{2}{c}{$T=1200$}\\
$m=20$&&${\min\atop{1\leq i\leq m}}EP_i$&$EP(\wh r=r)$&&${\min\atop{1\leq i\leq m}}EP_i$&$EP(\wh r=r)$&&${\min\atop{1\leq i\leq m}}EP_i$&$EP(\wh r=r)$\\
\cline{1-2}\cline{3-4}\cline{6-7}\cline{9-10}
$p_i=20$&&1&1&&1&1&&1&1\\
$p_i=30$&&1&1&&1&1&&1&1\\
$p_i=40$&&1&1&&1&1&&1&1\\
$p_i=50$&&1&1&&1&1&&1&1\\
\hline
\hline
\end{tabular}
  \end{center}
\end{table}

\begin{figure}
\begin{center}
\subfigure[]{\includegraphics[width=0.49\textwidth]{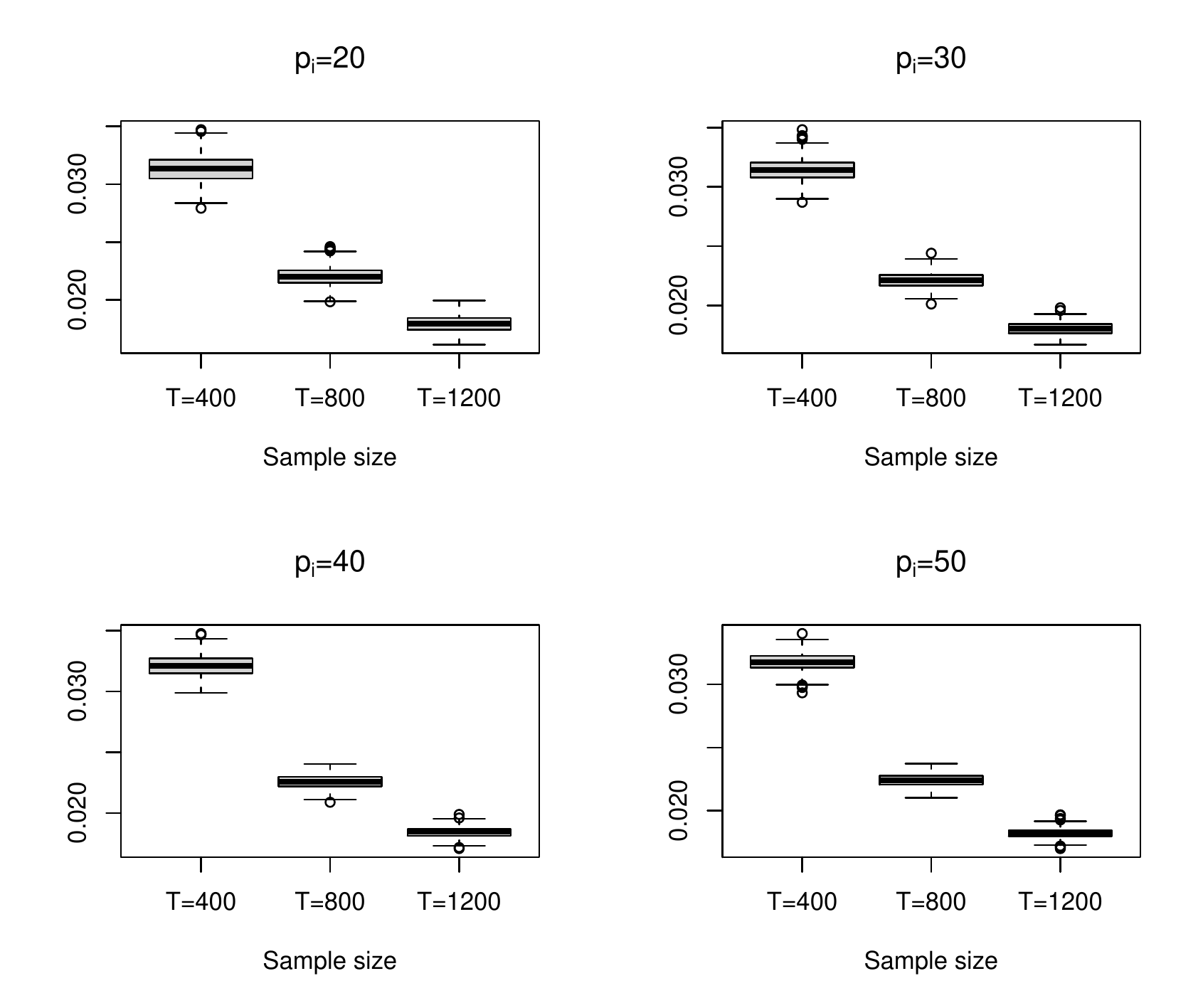}}
\subfigure[]{\includegraphics[width=0.49\textwidth]{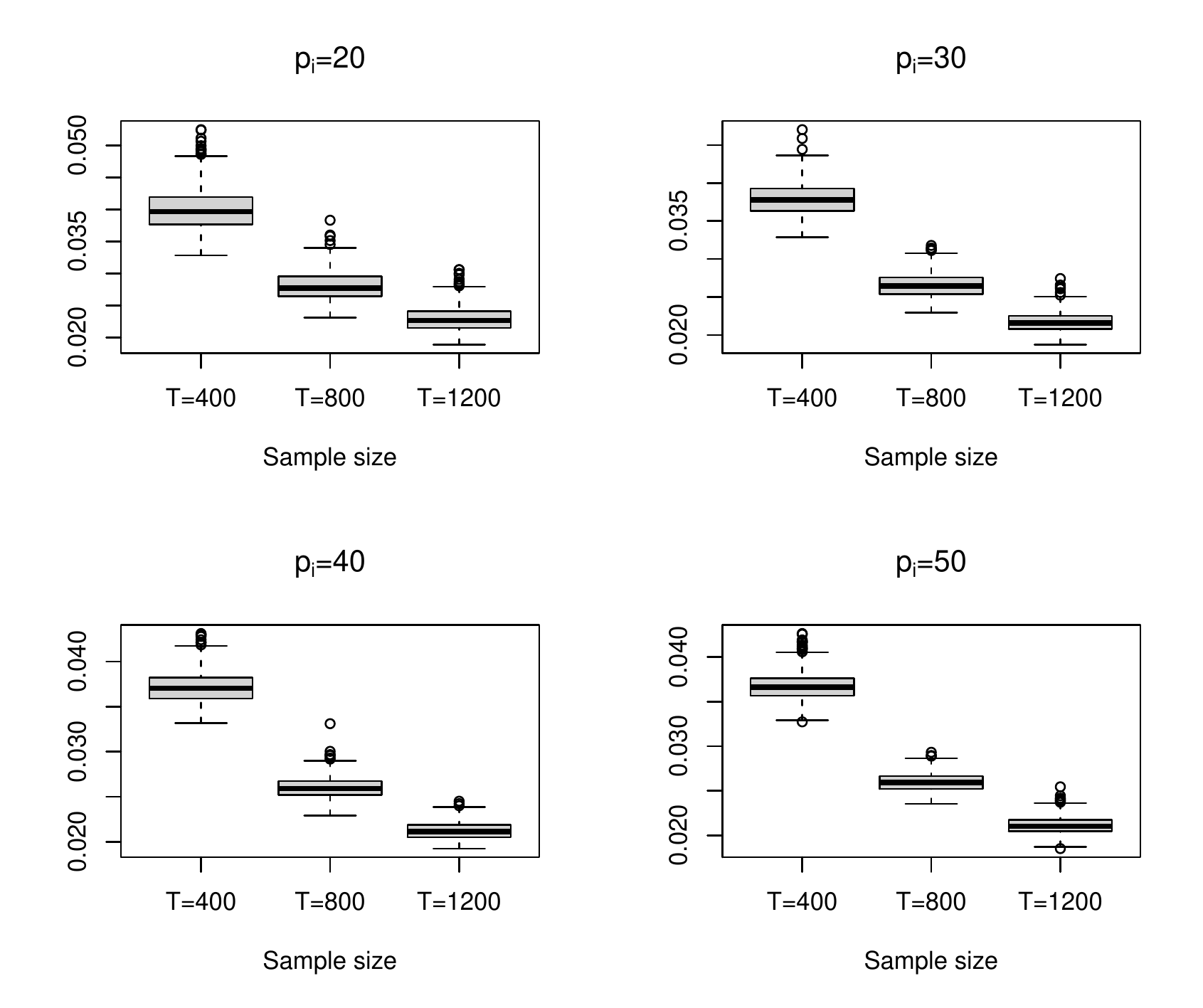}}
\caption{(a) Boxplots of the average discrepancy $\bar{D}_1(\wh\bA,\bA)$ when $r_i=5$ and $m=10$ of Example 1; (b) Boxplots of  the maximum discrepancy $\bar{D}_2(\wh\bA,\bA)$ when $r_i=5$ and $m=10$ of Example 1, where $\bar{D}_1(\cdot,\cdot)$ and  $\bar{D}_2(\cdot,\cdot)$ are defined in (\ref{da:am}). The sample sizes used are $400,\ 800,\ 1200$, and the results are based on $500$ iterations.}\label{fig1}
\end{center}
\end{figure}

\begin{figure}
\begin{center}
{\includegraphics[width=0.6\textwidth]{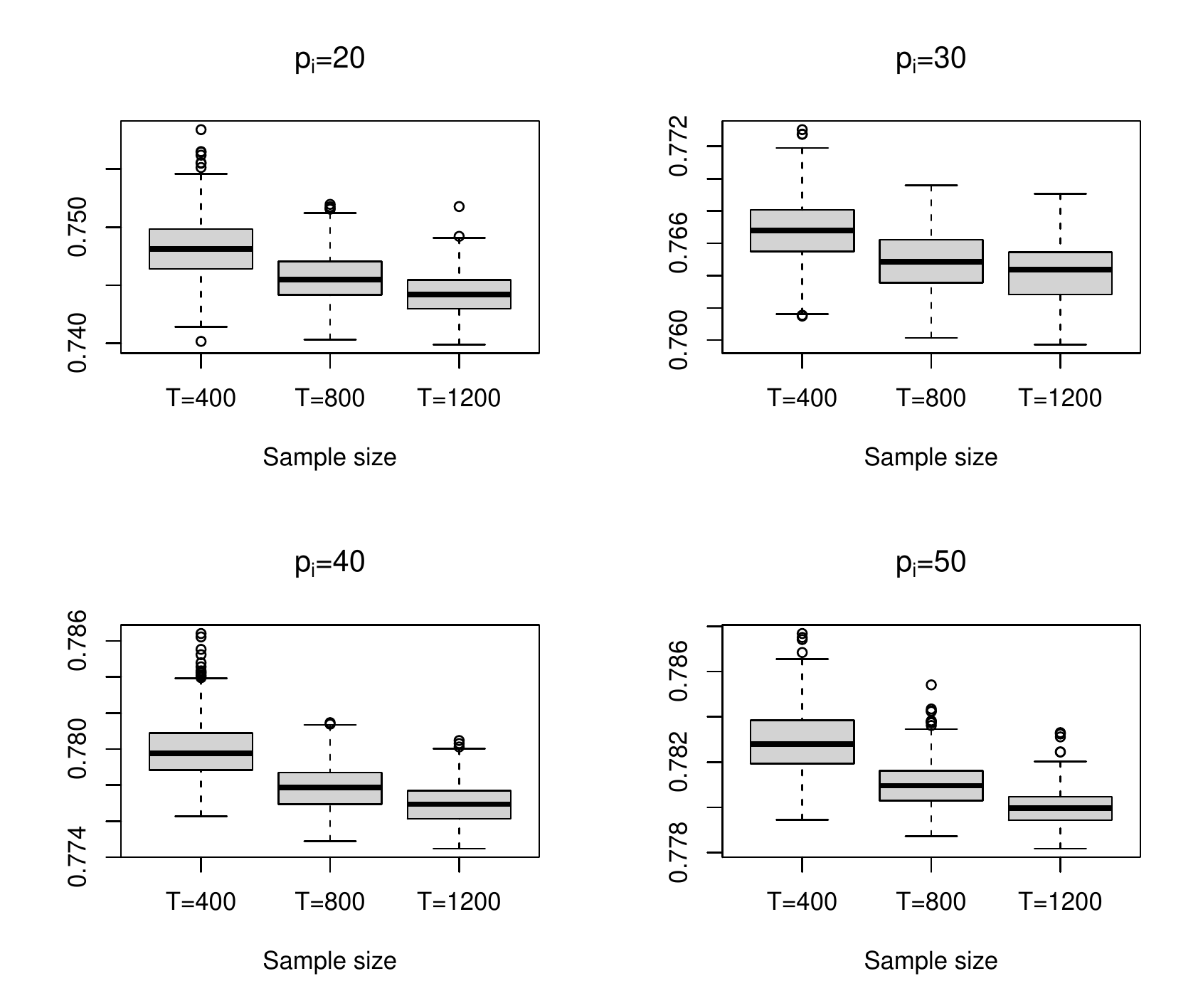}}
\caption{Boxplots of  $\bar{D}_3(\wh\bA\wh\bB,\bA\bB)$ with $r_i=5$, $r=3$, and $m=10$ of  Example 1, where $\bar{D}_3(\wh\bA\wh\bB,\bA\bB)$  is defined in (\ref{dab:m}). The sample sizes used are $T=400,\ 800,\ 1200$, 
and the number of iterations is $500$.}\label{fig2}
\end{center}
\end{figure}

Next, we study the estimation accuracy of the loading matrices. We only show the results for $m=10$  since the results are similar for $m$ = 20. Define
\begin{equation}\label{da:am}
\bar{D}_{1}(\wh\bA,\bA)=\frac{1}{m}\sum_{i=1}^m \bar{D}(\wh\bA_i\bA_i)\,\,\text{and}\,\,\bar{D}_2(\wh\bA,\bA)=\max_{1\leq i\leq m}\bar{D}(\wh\bA_i,\bA_i),
\end{equation}
which measure the average and the maximum  estimation errors between the linear spaces spanned by $\wh\bA_i$ and $\bA_i$, respectively. The boxplots of $\bar{D}_1(\wh\bA,\bA)$ and $\bar{D}_2(\wh\bA,\bA)$ are presented in Figure \ref{fig1}. From Figure \ref{fig1}(a)--(b), we see that, for each $p_i$, the discrepancy decreases  as the sample size increases and this is in agreement with our asymptotic theory. For the estimation errors of the loading matrix associated with the global factors, we further define
\begin{equation}\label{dab:m}
\bar{D}_3(\wh\bA\wh\bB,\bA\bB)=\frac{1}{m}\sum_{i=1}^m\bar{D}(\wh\bA_i\wh\bB_i,\bA_i\bB_i),
\end{equation}
which measures the average discrepancy between $\wh\bA_i\wh\bB_i$ and $\bA_i\bB_i$ in each group. The boxplots of $\bar{D}_3(\wh\bA\wh\bB,\bA\bB)$ are shown in Figure \ref{fig2}. From 
the figure, we also see that, for each $p_i$, the discrepancy between the estimated global loading matrix and the true ones decreases as the sample size increases  and the result is also in agreement with our theorems.

\begin{figure}
\begin{center}
\subfigure[]{\includegraphics[width=0.49\textwidth]{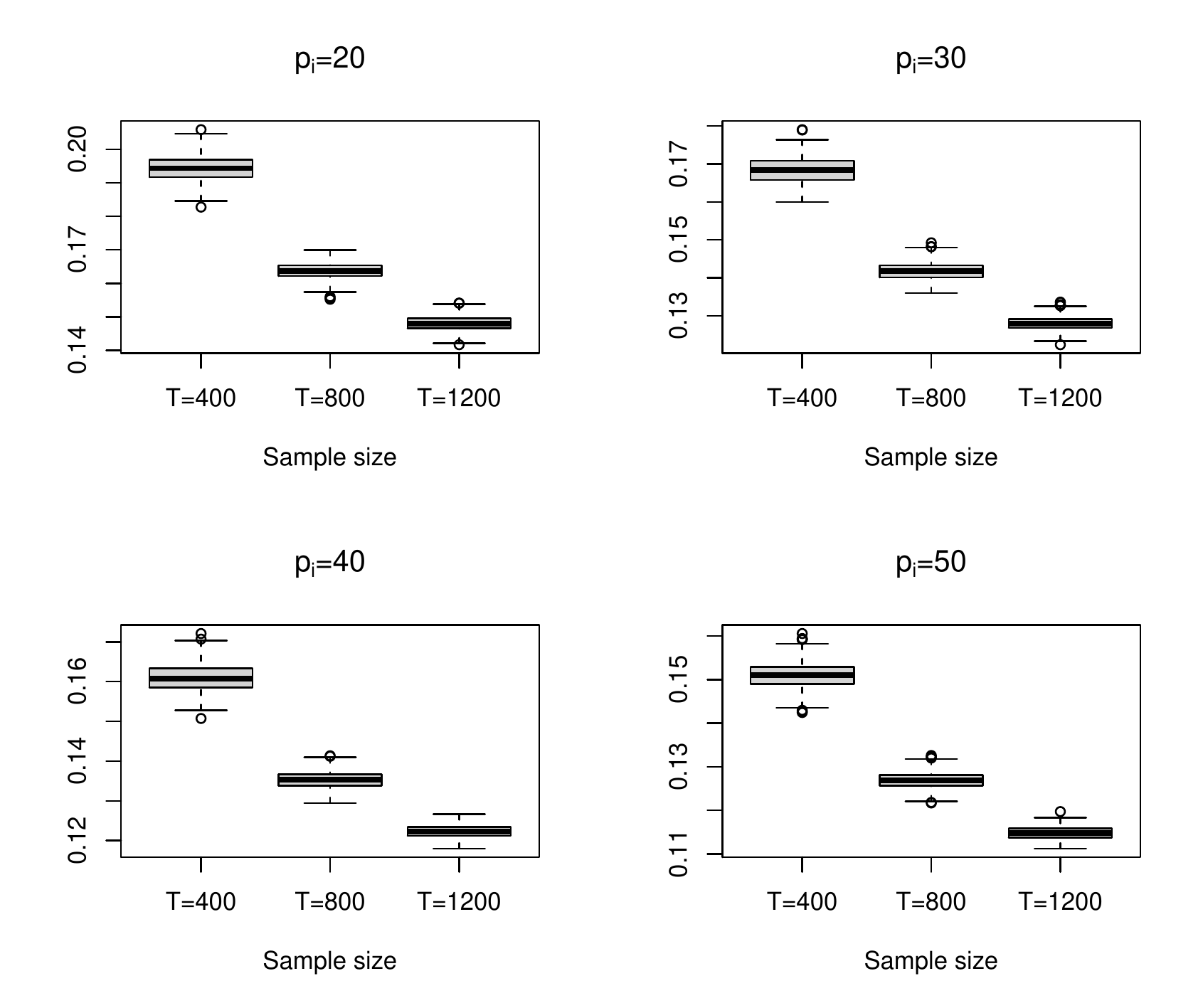}}
\subfigure[]{\includegraphics[width=0.49\textwidth]{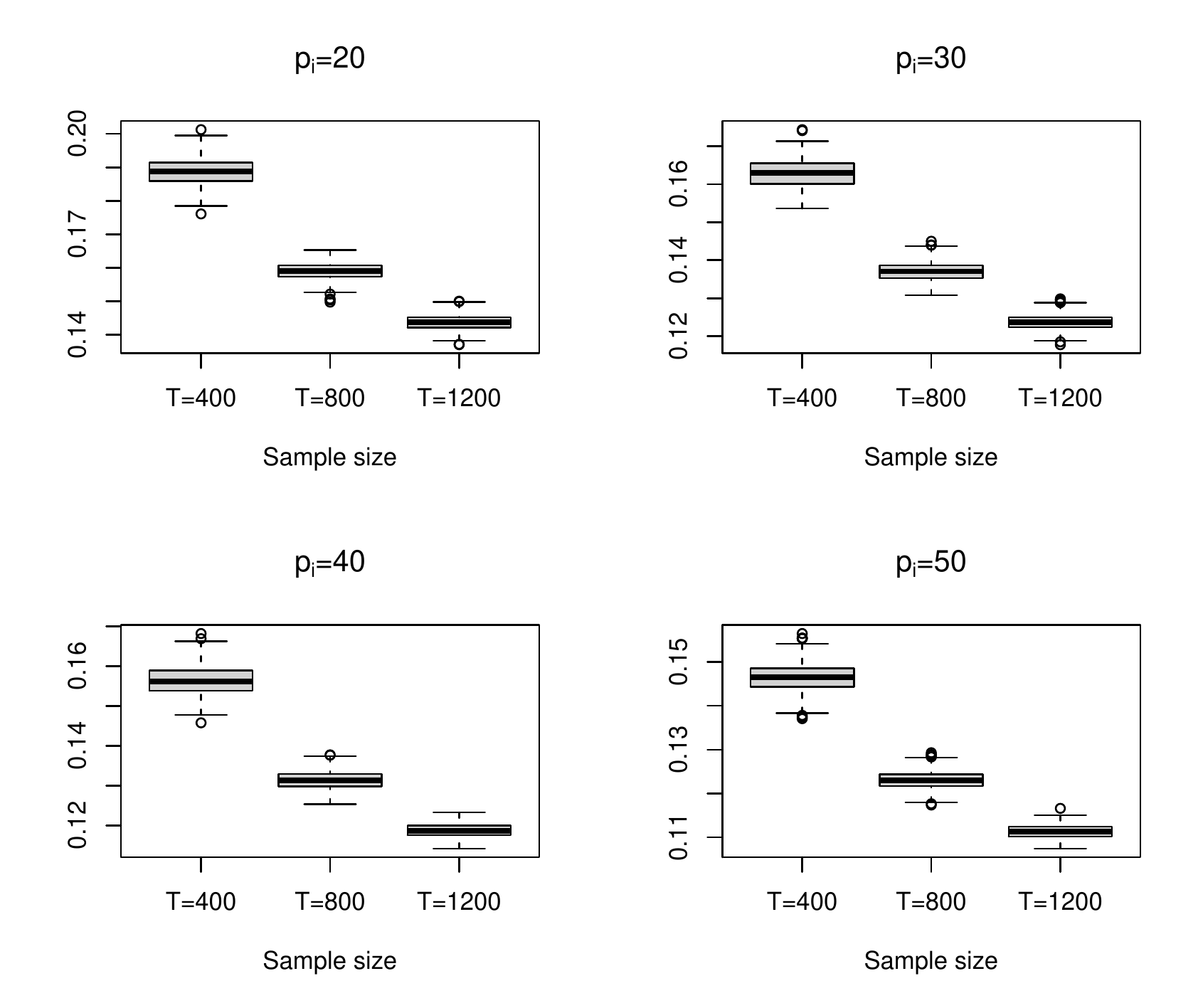}}
\caption{(a) Boxplots of RMSE$_1$ with $r_i=5$ and $m=10$ in Example 1; (b) Boxplots of  RMSE$_2$ with $r_i=5$ and $m=10$ in Example 1. The sample sizes used are $400,\ 800,\ 1200$, and the results are based on $500$ iterations.}\label{fig3}
\end{center}
\end{figure}

Finally, we study the estimation errors of the extracted factors. We only consider the case when $m=10$ since the results are similar for $m$ = 20. For each $(p_i,T)$, define the root-mean-squared error (RMSE) as 
\begin{equation}\label{rmse:f}
\text{RMSE}_1=(\frac{1}{p_imT}\sum_{t=1}^T\|\wh\bA\wh\bgg_t-\bA\bgg_t\|_2^2)^{1/2},\,\,\text{RMSE}_2=(\frac{1}{p_imT}\sum_{t=1}^T\|\wh\bA\wh\bB\wh\bff_t-\bA\bB\bff_t\|_2^2)^{1/2},
\end{equation}
which quantify the accuracy in estimating the group factors and the global factors, respectively. Boxplots of the RMSEs are shown in Figure \ref{fig3}. From the plot, we see clearly that, as the sample size increases, the RMSE decreases for a given $p_i$, which is consistent with the results of Theorem \ref{tm3}.
Also, as expected from the same theorem, the RMSE decreases with the dimension $p_i$. Overall, our proposed method works well for various settings of moderately large dimensions and sample sizes.

\begin{figure}
\begin{center}
\subfigure[]{\includegraphics[width=0.49\textwidth]{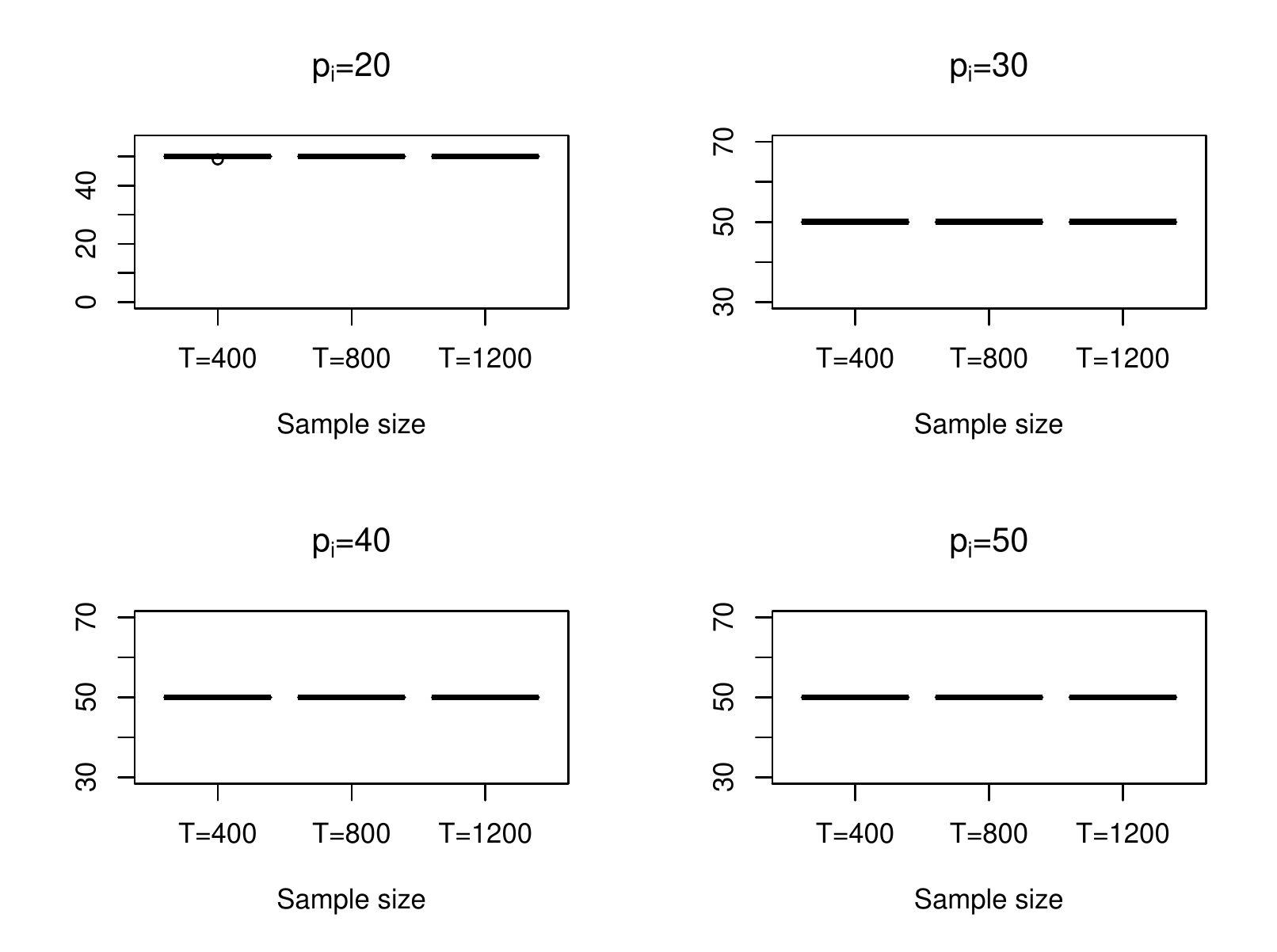}}
\subfigure[]{\includegraphics[width=0.49\textwidth]{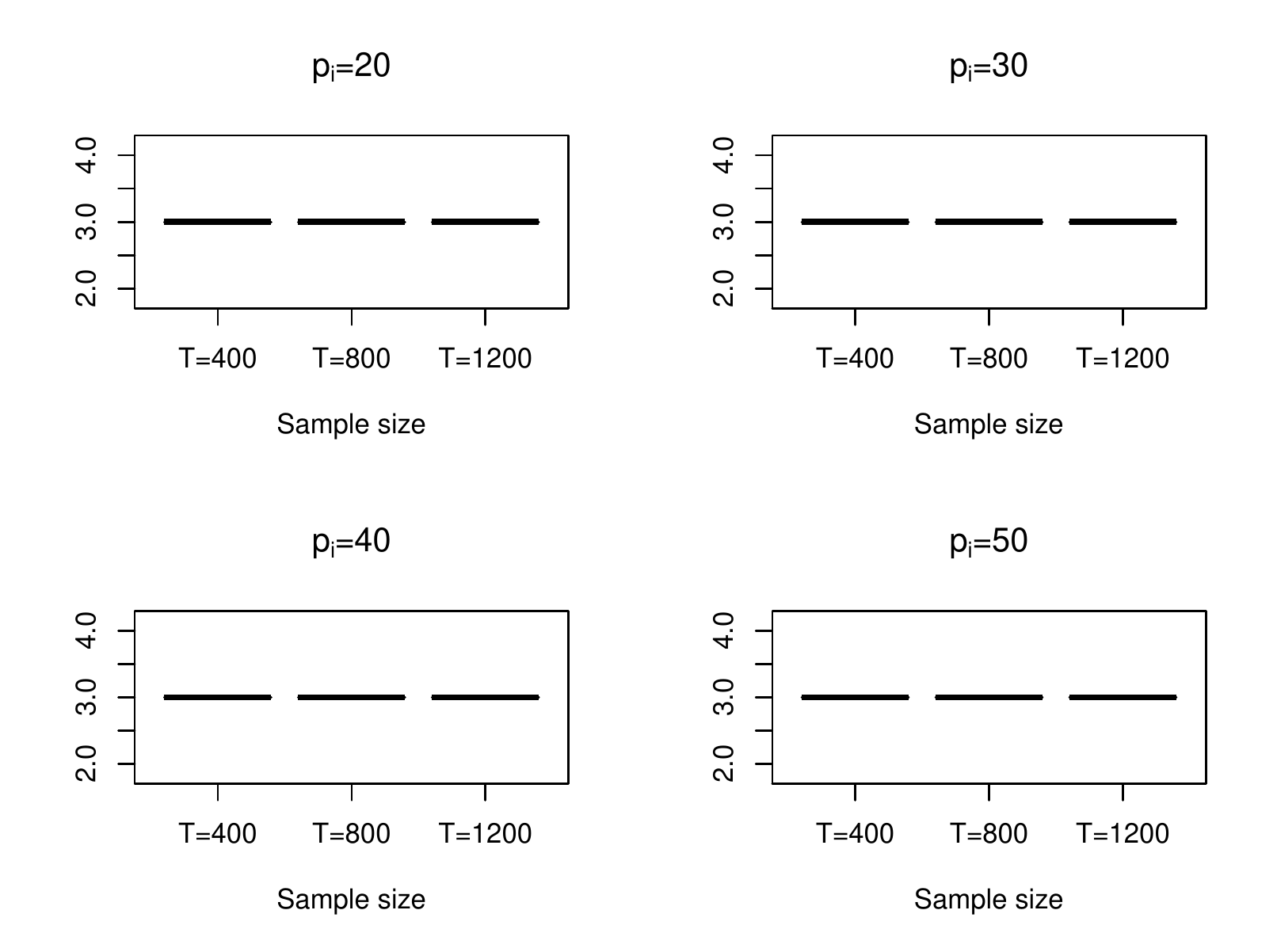}}
\caption{(a) Boxplots of the estimated number of factors using the full data by the IC1 criterion in \cite{BaiNg_Econometrica_2002} when $r_i=5$, $r=3$, and $m=10$ in Example 1; (b) Boxplots of  the estimated number of factors using the full data by the ratio-based method in \cite{ahn2013} and \cite{lamyao2012} when $r_i=5$, $r=3$, and $m=10$ in Example 1. The sample sizes used are $400,\ 800,\ 1200$, and the results are based on $500$ iterations.}\label{fig-r}
\end{center}
\end{figure}

Next, we apply the BIC-type criterion IC1 of Section 2.2 as suggested by \cite{BaiNg_Econometrica_2002} and the ratio-based method in \cite{ahn2013} and \cite{lamyao2012} to the entire data set. Our goal here is to illustrate the impact of overlooking 
the underlying factor structure of the model in data analysis.  The boxplots of the estimated numbers of common factors by these two methods are shown in Figure~\ref{fig-r}. From the plots, we see that, 
overlooking the hierarchical factor structure, the BIC criterion of \cite{BaiNg_Econometrica_2002} is able to identify the total number of the group factors while ignoring the number of global common factors, and the ratio-based method tends to identify the number of global factors while ignoring the total number of group factors. On the other hand, as shown in Table \ref{Table1}, our hierarchical framework works well to identify both numbers of common factors.

\vskip 0.5cm

\noindent{\bf Example 2.} In this example, we consider the models in (\ref{m-factor}) and (\ref{global-f}) with global common factors following the random walk model
\[\bff_t=\bff_{t-1}+\bfeta_t,\]
where $\bfeta_t$ is a white noise process. We set $r_i=3$ and $r=5$, and everything else is generated in the same way as that of Example 1. 

To study the performance of estimating the number of stochastic trends, we use the the autocorrelation-based method in (\ref{auto:cor}) with $\delta_0=0.3$ and $\bar{k}=20$ to estimate $r_i$ and $r$. The empirical probabilities of $\min_{1\leq i\leq m}P(\wh r_i=r_i)$ and $P(\wh r=r)$ are reported in Table~\ref{Table2-1} for $m=10$. Similar results are also found for the other choice of $m$, and we omit the detail to save space. From Table~\ref{Table2-1}, we see that the proposed method works well in selecting the numbers of group common stochastic trends and the global ones. 
Overall, as the sample size  increases, the proposed method shows improved performance in estimating the numbers of common factors, which is in agreement with the asymptotic theory. 
In addition, the performance of the proposed method seems to be stable or even improved 
as the dimension increases, showing evidence of the `blessing of dimensionality' in high-dimensional PCA estimation. The boxplots of the empirical discrepancies between the estimated loading 
matrices and the true ones are similar to those in Example 1 because the estimated numbers of factors are quite accurate according to Table~\ref{Table2-1}. Therefore, we do not repeat them here.


\begin{table}
 \caption{Empirical probabilities $\min_{1\leq i\leq m}EP(\hat{r}_i=r_i)$, 
 denoted by $\min_{1\leq i\leq m}EP_i$, and $EP(\wh r=r)$ of various $(p_i,T)$ configurations 
 for the model of Example 2 with $m=10$, $r_i=3$ and $r=5$, where $p_i$ and $T$ are, respectively,  the dimension and the sample size of each group. $500$ iterations are used.} 
          \label{Table2-1}
\begin{center}
 \setlength{\abovecaptionskip}{0pt}
\setlength{\belowcaptionskip}{3pt}

\begin{tabular}{c|ccccccccc}
\hline
\hline
&&\multicolumn{2}{c}{$T=400$}&&\multicolumn{2}{c}{$T=800$}&&\multicolumn{2}{c}{$T=1200$}\\
$m=10$&&${\min\atop{1\leq i\leq m}}EP_i$&$EP(\wh r=r)$&&${\min\atop{1\leq i\leq m}}EP_i$&$EP(\wh r=r)$&&${\min\atop{1\leq i\leq m}}EP_i$&$EP(\wh r=r)$\\
\cline{1-2}\cline{3-4}\cline{6-7}\cline{9-10}
$p_i=20$&&0.982&0.980&&1&1&&1&1\\
$p_i=30$&&0.984&0.984&&1&1&&1&1\\
$p_i=40$&&0.992&0.982&&1&1&&1&1\\
$p_i=50$&&0.996&0.994&&1&1&&1&1\\
\hline
\hline
\end{tabular}
  \end{center}
\end{table}

\subsection{Real Data Analysis}
In this section, we demonstrate the application of the proposed method using U.S. monthly macroeconomic data with 157 time series and 480 observations. 

\noindent{\bf Example 3.} Consider the data set of U.S. monthly macroeconomic time series available from January 1959 to December 1998, which was used in \cite{StockWatson_2002a}. The description of the original data set consisting of 215 series was given in \cite{StockWatson_2002a}, but we eliminate the series with missing data and the series in the categories of average hourly earnings and   miscellaneous, because all, but one or two, series in these two categories contain missing 
values. The resulting data set consists of 157 series with $480$ observations. To achieve stationarity, we take the first difference of all series and, hence, we employ the transformed data 
with $p=157$ and $T=479$.

\begin{figure}
\begin{center}
{\includegraphics[width=0.8\textwidth]{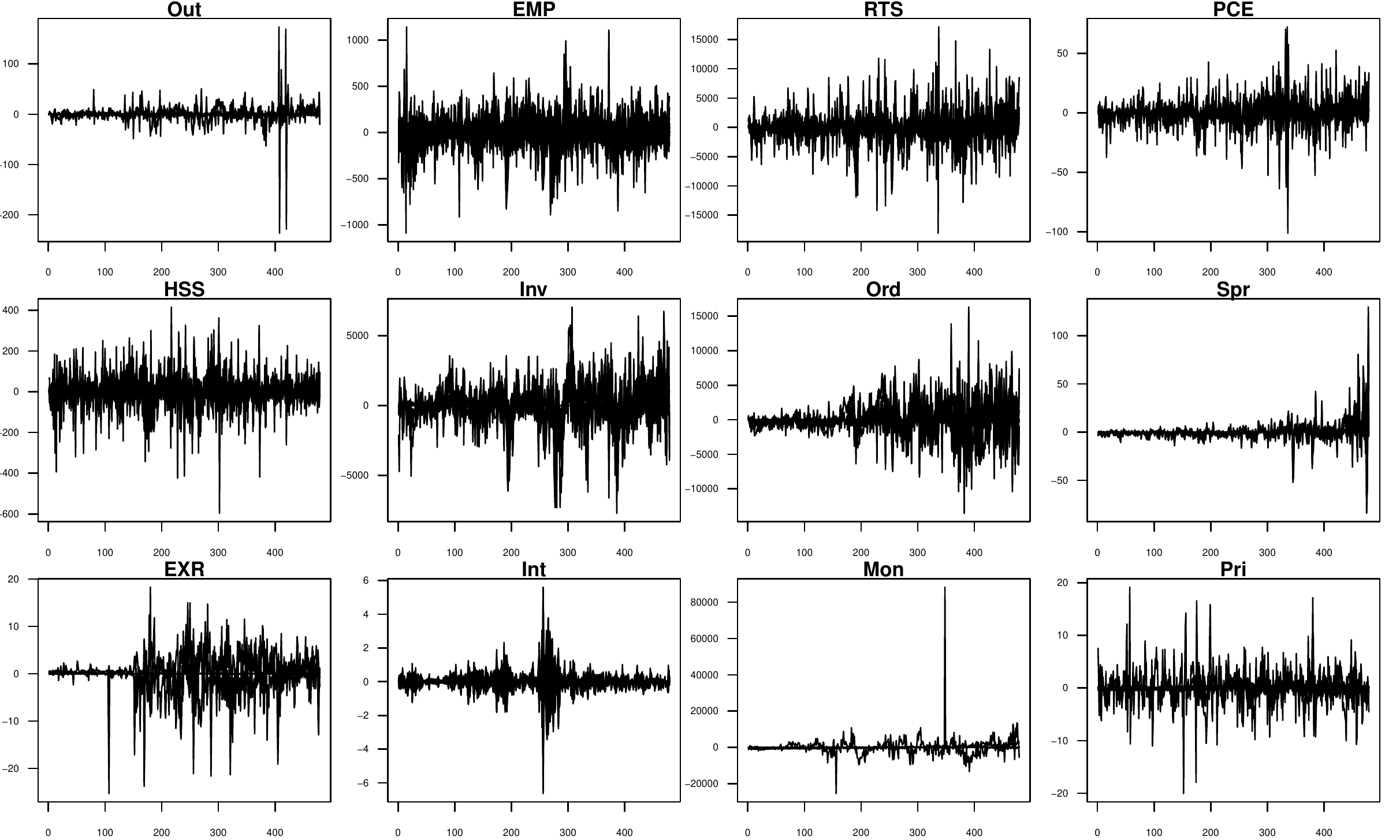}}
\caption{Time plots of the differenced series in 12 categories (or groups) from February 1959 to December 1998. 
They are (1) Real output and income (Out), (2) Employment and hours (EMP), (3) Retail and manufacturing trade (RTS), (4) Consumption (PCE), (5) Housing starts and sales (HSS), (6) Real inventories and inventory-sales ratios (Inv), (7) Orders and unfilled orders (Ord), (8) Stock prices (Spr), (9) exchange rates (EXR),  (10)  Interest rates (Int),   (11) Money and credit quantity aggregates (Mon), (12) Price indexes (Pri). }\label{fig4}
\end{center}
\end{figure}

The series can roughly be classified into $12$ groups (or clusters) as indicated in \cite{StockWatson_2002a}: (1) Real output and income (Out, series 1--21), (2) Employment and hours (EMP, series 22--48), (3) Retails and manufacturing trade (RTS, series 49--57), (4) Consumption (PCE, series 58--62), (5) Housing starts and sales (HSS, 53--69), (6) Real inventories and inventory-sales ratios (Inv, series 70--80), (7) Orders and unfilled orders (Ord, series 81--96), (8) Stock prices (Spr, series 97--103), (9) exchange rates (EXR, series 104--109),  (10)  Interest rates (Int, series 110--128),   (11) Money and credit quantity aggregates (Mon, series 129--137), (12) Price indexes (Pri, series 138--157). Details are given in \cite{StockWatson_2002a} and the series of 12 groups are shown in Figure \ref{fig4} with $m=12$.

\begin{figure}
\begin{center}
{\includegraphics[width=0.85\textwidth]{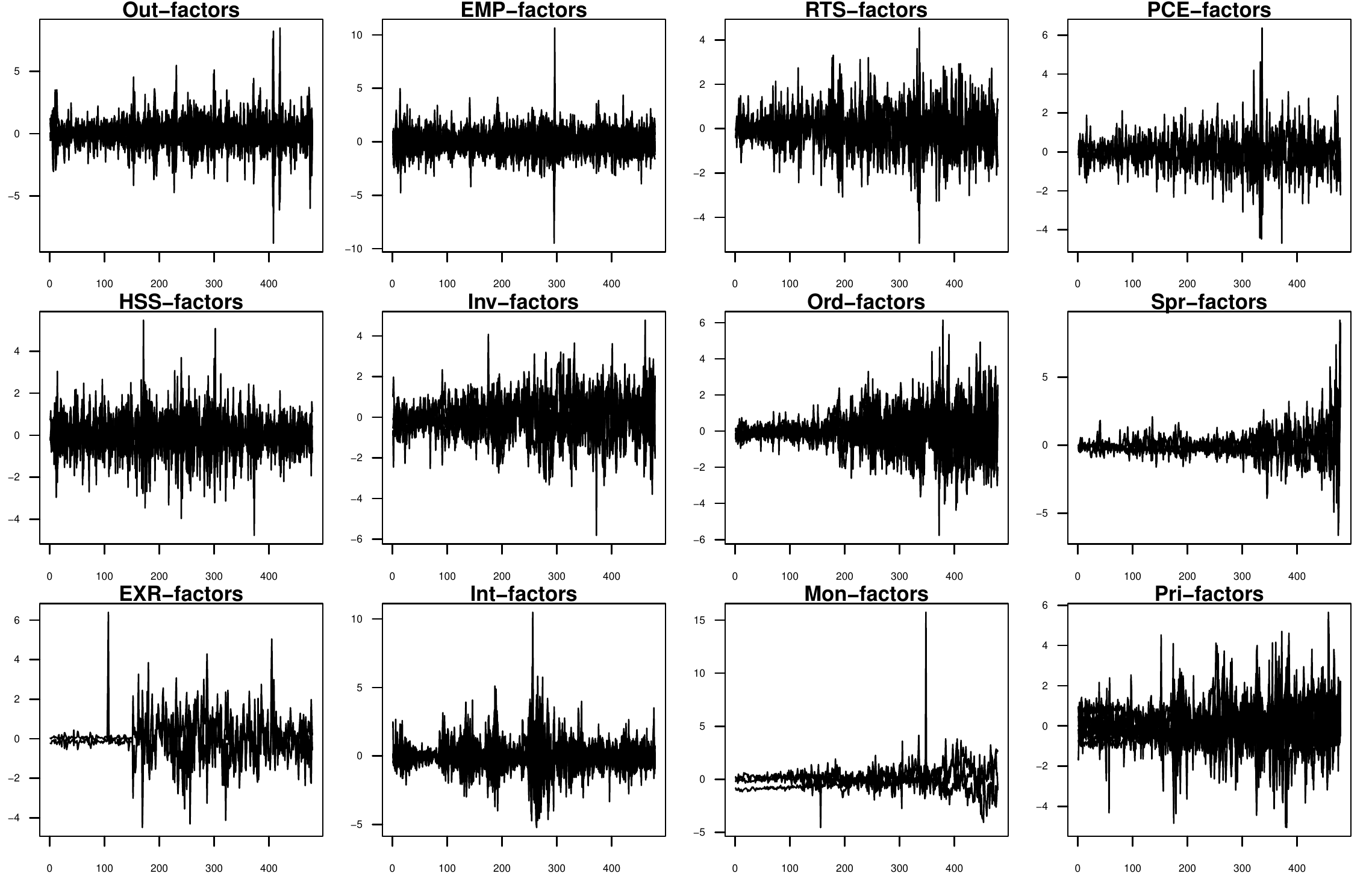}}
\caption{Time series plots of group common factors of 12 categories from February 1959 to December  1998. They are (1) Real output and income (Out, $r_1=7$), (2) Employment and hours (EMP, $r_2=9$), (3) Retail and manufacturing trade (RTS, $r_3=3$), (4) Consumption (PCE,  $r_4=2$), (5) Housing starts and sales (HSS, $r_5=3$), (6) Real inventories and inventory-sales ratios (Inv, $r_6=4$), (7) Orders and unfilled orders (Ord, $r_7=6$ ), (8) Stock prices (Spr,  $r_8=3$), (9) exchange rates (EXR, $r_9=2$,),  (10)  Interest rates (Int, $r_{10}=7$),   (11) Money and credit quantity aggregates (Mon, $r_{11}=3$), (12) Price indexes (Pri, $r_{12}=7$). }\label{fig5}
\end{center}
\end{figure}

\begin{figure}
\begin{center}
{\includegraphics[width=0.9\textwidth]{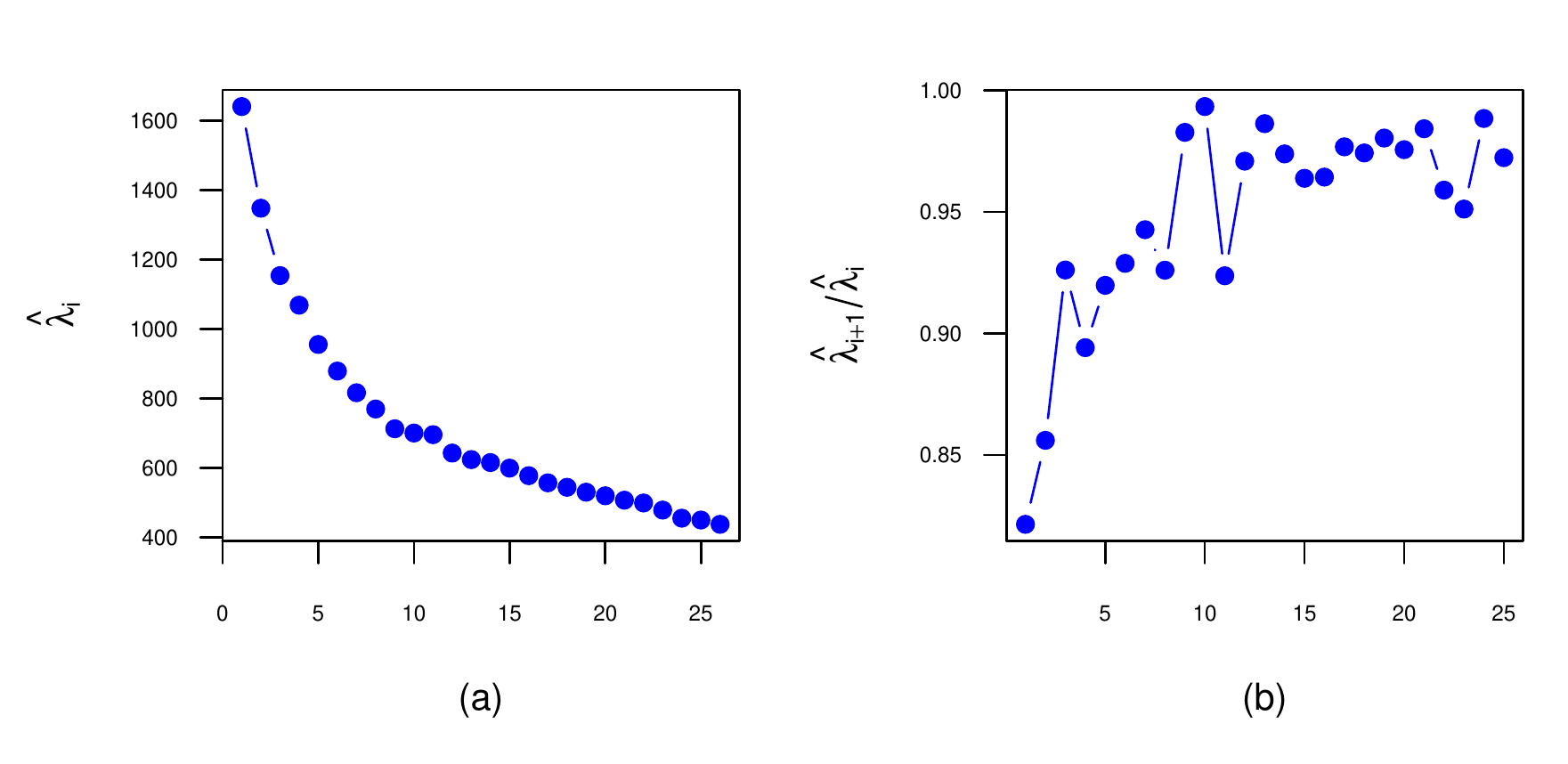}}
\caption{(a) The first 26 eigenvalues of $\wh\bG'\wh\bG$; (b) Ratios of consecutive eigenvalues
of $\wh\bG'\wh\bG$ in Example 3.}\label{fig6}
\end{center}
\end{figure}

\begin{figure}
\begin{center}
{\includegraphics[width=0.9\textwidth]{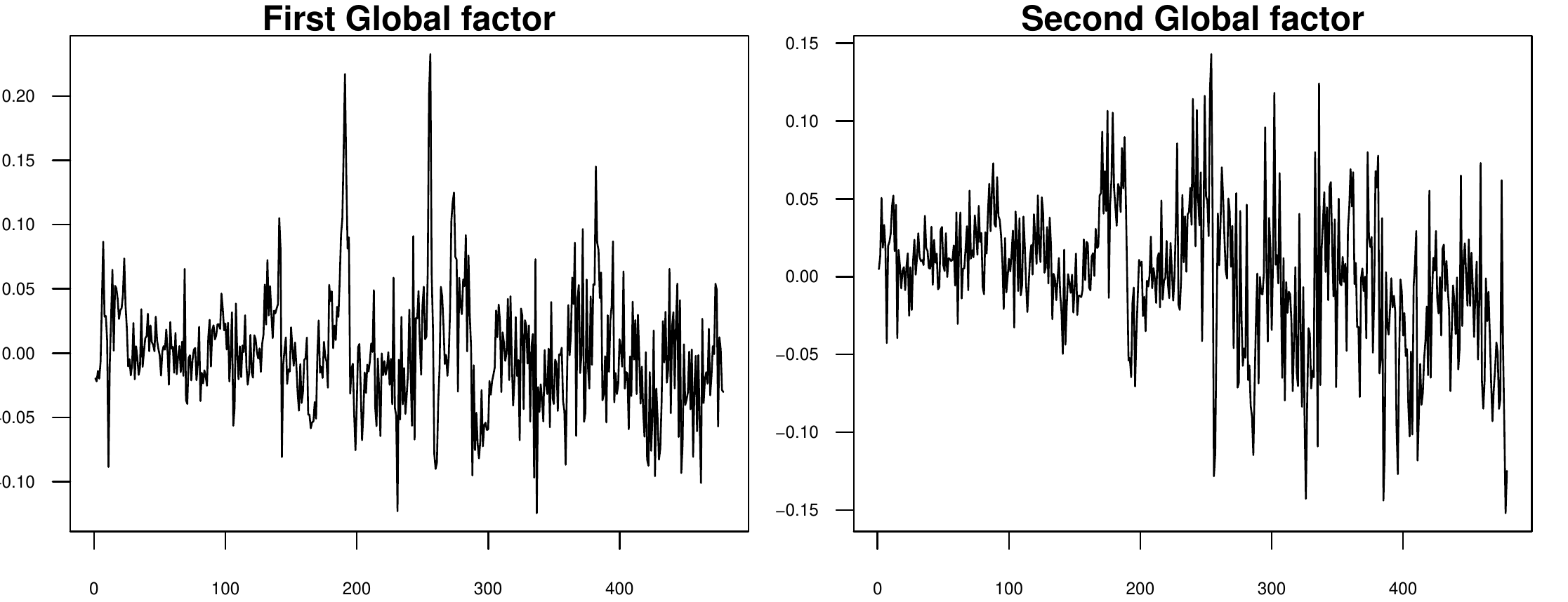}}
\caption{(a) Time series plot of the first global factor process $\wh f_{1,t}$  of Example 3; (b) Time series plot of the second global factor process $\wh f_{2,t}$ of Example 3.}\label{fig7}
\end{center}
\end{figure}

We first applied our proposed method to each group and found that the numbers of 
common factors in each group are (1) $r_1=7$, (2) $r_2=9$, (3) $r_3=3$, (4) $r_4=2$, (5) $r_5=3$, (6) $r_6=4$, (7) $r_7=6$, (8) $r_8=3$, (9) $r_9=2$, (10) $r_{10}=7$, (11) $r_{11}=3$, and (12) $r_{12}=7$. 
The total number of group factors is $k_m=56$. The time plots of all group factors are shown in Figure~\ref{fig5}. We then performed the second stage factor analysis. The first $26$ eigenvalues ($\wh\lambda_i$) of $\wh\bG'\wh\bG$ and the associated eigenvalue ratios are presented in Figure~\ref{fig6}. From the plots, we see that the sharpest drop occurs at the first ratio, which implies that there is one global common factor and, hence, $\wh r=1$. For ease in illustration, we show the first two estimated global factors in Figure~\ref{fig7}. These factors capture the common cross-sectional  dependence between the group factors of series in Figure~\ref{fig5}.

\begin{figure}
\begin{center}
{\includegraphics[width=0.9\textwidth]{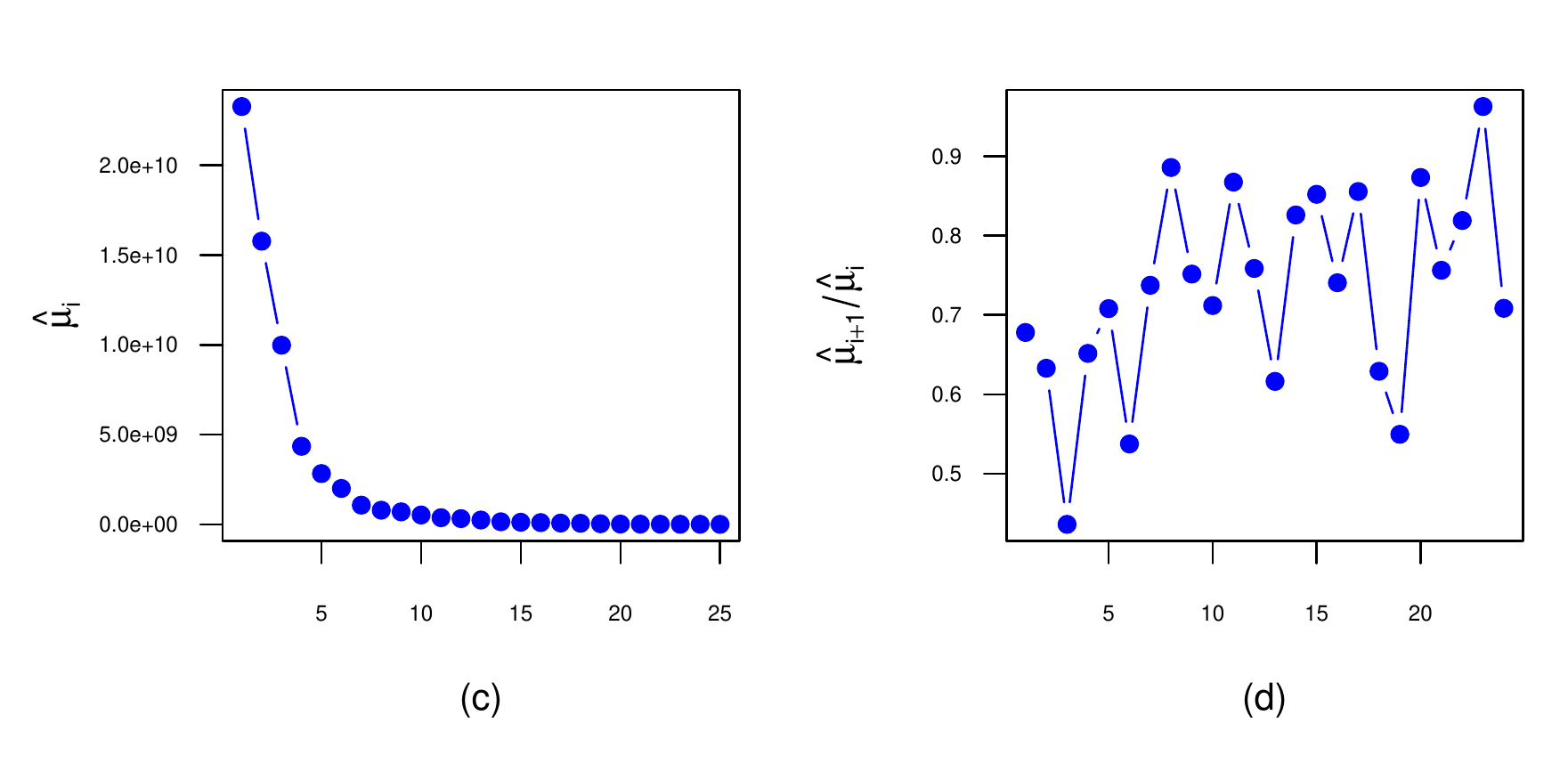}}
\caption{(a) The first 25 eigenvalues of $\wh\bSigma_y$; (b) The plot of ratios of consecutive eigenvalues
of $\wh\bSigma_y$ in Example 3.}\label{fig8}
\end{center}
\end{figure}

Next, we compare the forecasting ability of our global and group-specific factors with the ones obtained using the whole $157$ series jointly. Observations of the last ten years are used for out-of-sample forecasting. When using the information criterion method of \cite{BaiNg_Econometrica_2002} to select the number of factors, we find that $\wh r$ increases as we increase the upper bound $K$ in (\ref{ri}), and it eventually reaches a value greater than $50$, so we adopt the ratio-based method in \cite{ahn2013} and \cite{lamyao2012}, as specified in a similar fashion as (\ref{est-r}). The first 25 eigenvalues, $\wh\mu_i$, of the covariance of the full data $\wh\bSigma_y$ and the ratios of the consecutive eigenvalues are given in Figure~\ref{fig8}.  The sharpest drop of the ratios occurs at $\wh r_{full}=3$. For comparisons of the forecasting ability of different factors, we investigate the 
1-step to 4-step  ahead forecasts of eight commonly studied economic indicators ($x_{i,t}, 1\leq i\leq 8$). They are the industrial production (ip), real personal income less transfers (gmyxpq), real manufacturing trade and sales (msmtq), number of employees on non-agricultural payrolls (lpnag), the consumer price index (CPI) (punew), the personal consumption expenditure deflator (gmdc), the consumer price index less food and energy (puxf), and the producer price index for finished goods (pwfsa); see, also, \cite{StockWatson_2002a}.

To see the forecastability of the extracted factors, we only regress $x_{i,t+h}$ on $\wh\bff_t$ and $\wh\bu_{i,t}$ without including the lagged variable of $x_{i,t}$, because \cite{StockWatson_2002a} found that the simple diffusion index forecasts excluding the lags are even better than that with the lagged variables. Table \ref{Table2} reports the $1$-step to $4$-step ahead mean squared forecast errors for the eight selected series described above. We adopt five models according to the included factors in (\ref{df-id}). They are (1) 1glb: including the first global factor in Figure~\ref{fig7}(a) as a predictor; (2) 2glb: including the two global factors in Figure~\ref{fig7}(a,b) as predictors; 
(3) (1glb+spf): including the 
first global factor and the associated group-specific factors $\bu_{i,t}$ of $x_{i,t}$; 
(4) 2glb+spf: including the two global factors and the associated group-specific factors $\bu_{i,t}$ of $x_{i,t}$ (2glb+spf); (5) SW: including the 3-dimensional factor process obtained from the full data set as that in \cite{StockWatson_2002a}. From Table \ref{Table2}, we see that the smallest error was never achieved by the SW method  except the 2-step ahead prediction for gmyxpq. In addition, the forecastability of the first global factor identified by our ratio-based method performs better than that of the factors recovered using the full data set by the method of \cite{StockWatson_2002a}, except for the 1-step ahead prediction of msmtq; the method of 2glb outperforms SW except for the 1-step ahead forecast of msmtq, and the 2-step ahead forecast for gmyxpq; the method of 1glb+spf outperforms SW except for the 1-step ahead forecast for gmyxpq, the 2-step ahead forecasts of gmyxpq and pwfsa, and the 4-step ahead forecasts for ip, gmyxpq, and pwfsa; the method of 2glb+spf outperforms SW except for the 1-step ahead forecast of gmyxpq, the 2-step ahead forecast for gmyxpq and pwfsa, and the 4-step ahead forecasts for ip and pwfsa. Overall, the factors extracted by the proposed method tend to have higher forecastability than those obtained by the full data set in predicting the 
selected eight economic variables.


\begin{table}[h]
\footnotesize{
 \caption{The 1-step to 4-step ahead forecast errors. Five methods employed are according to 
 factors used in (\ref{df-id}). They are (1) 1glb: including the first global factor in Figure~\ref{fig7}(a) as a predictor; (2) 2glb: including two global factors in Figure~\ref{fig7}(a,b) as predictors; (3)
 1glb+spf: including 
 the frist global factor and the associated group-specific factors $\bu_{i,t}$ of $x_{i,t}$; (4) 
 2glb+spf: including two global factors and the associated group-specific factors $\bu_{i,t}$ of $x_{i,t}$; (5) SW: including the 3-dimensional factor process obtained from the full data as that in \cite{StockWatson_2002a}.
 Boldface numbers denote the smallest one for a given variable and step. In the table, msmtq$^*$ and 
 lpnag$*$ denote, respectively, msmtq$/10^7$ and lpnag$/10^4$.} 
          \label{Table2}
\begin{center}
 \setlength{\abovecaptionskip}{0pt}
\setlength{\belowcaptionskip}{3pt}

\begin{tabular}{c|c|cccccccc}
\hline
\hline
Step&Method&ip&gmyxpq&msmtq$^*$ &lpnag$^*$ &punew&gmdx&puxf&pwfsa\\
\hline
1&1glb&0.315&{\bf 268.7}&2.8&2.1&0.050&0.024&0.054&0.212\\
&2glb&0.322&270.9&2.9&2.3&0.050&0.024&0.054&{\bf 0.194}\\
&1glb+spf&{\bf 0.278}&281.2&2.7&{\bf 1.7}&{\bf 0.040}&{\bf 0.020}&{\bf 0.045}&0.198\\
&2glb+spf&0.284&283.8&{\bf 2.6}&{\bf 1.7}&{\bf 0.040}&{\bf 0.020}&{\bf 0.045}&0.200\\
&SW&0.346&279.0&2.7&2.8&0.052&0.024&0.055&0.201\\
\hline
2&1glb&0.305&272.8&2.8&2.2&0.051&0.024&0.054&0.212\\
&2glb&0.285&273.6&2.6&2.2&0.051&0.024&0.056&{\bf 0.203}\\
&1glb+spf&0.301&291.9&2.7&{\bf 1.6}&0.050&{\bf 0.023}&{\bf 0.050}&0.228\\
&2glb+spf&{\bf 0.290}&292.8&{\bf 2.5}&{\bf 1.6}&{\bf 0.049}&{\bf 0.023}&{\bf 0.050}&0.225\\
&SW&0.338&{\bf 272.1}&2.8&2.4&0.058&0.027&0.062&0.218\\
\hline
3&1glb&0.286&272.9&{\bf 2.6}&1.8&0.051&0.024&0.055&0.212\\
&2glb&{\bf 0.269}&262.9&{\bf 2.6}&1.7&0.053&0.024&0.057&{\bf 0.209}\\
&1glb+spf&0.305&268.6&2.7&{\bf 1.4}&0.044&{\bf 0.020}&0.048&0.215\\
&2glb+spf&0.290&{\bf 256.8}&{\bf 2.6}&{\bf 1.4}&{\bf 0.043}&{\bf 0.020}&{\bf 0.046}&0.213\\
&SW&0.333&284.0&2.7&2.5&0.059&0.027&0.065&0.215\\
\hline
4&1glb&0.322&260.2&2.8&2.3&0.051&0.025&0.055&{\bf 0.212}\\
&2glb&{\bf 0.315}&{\bf 259.5}&{\bf 2.7}&2.3&0.052&0.025&0.056&0.215\\
&1glb+spf&0.341&267.2&2.8&{\bf 1.8}&0.046&{\bf 0.021}&{\bf 0.049}&0.240\\
&2glb+spf&0.333&262.6&2.8&{\bf 1.8}&{\bf 0.045}&{\bf 0.021}&0.050&0.236\\
&SW&0.332&263.5&2.8&2.7&0.060&0.027&0.064&0.220\\
\hline
\hline
\end{tabular}
  \end{center}}
\end{table}

\section{Concluding Remarks}\label{sec7}
This article introduced a distributed factor approach to modeling large-scale stationary and 
unit-root non-stationary time series data. In the modern era of data science, large data sets are usually scattered across different locations such that it is extremely difficult to fuse or aggregate them due to the issues of communication cost, privacy concern, data security, and ownership, among others. The proposed divide-and-conquer method can overcome this difficulty by performing the factor modeling in parallel, and the resulting factors are of lower dimensions and easier to transfer, and the privacy can also be protected. The numerical results suggest that the factors produced by our proposed method have higher forecasting ability than those obtained by applying the traditional PCA to the full data set. The proposed method thus provides another option in the toolkit for the practitioners who are interested in out-of-sample forecasting with large-scale data sets. We also briefly discussed 
the analysis of 
extremely large-scale time series data for which both the horizontal and vertical partitions are needed. 
This topic deserves a careful further investigation.

\section*{Appendix: Proofs}
\renewcommand{\theequation}{A.\arabic{equation}}
\setcounter{equation}{0}
In this section, we use $C$ or $c$ as a generic constant the value of which may change at different places.  Recall that $\bar{p}=\max\{p_1,...,p_m\}$ and $\underline{p}=\min\{p_1,...,p_m\}$.

{\bf Proof of Theorem 1.} Let $\bH_i=\wh\bV_i^{-1}(\wh\bG_i'\bG_i/T)(\bA_i'\bA_i/p_i)$, where $\wh\bV_i$ is a $r_i\times r_i$ diagonal matrix consisting of the $r_i$ largest eigenvalues of $\bY_i\bY_i'/(p_iT)$.  (\ref{rate:g}) follows from Theorem 1 of \cite{BaiNg_Econometrica_2002}, and  (\ref{max:rate1}) follows from Theorem 4 of \cite{fan2013}. We only prove the results in (\ref{rate:f}) and (\ref{max:rate2}).

Define $\bK'=\wh\bV_g^{-1}(\wh\bF\bF/T)(\bB'\bB/k_m)$ and $\bH=\diag(\bH_1,...,\bH_m)$, where $\wh\bV_g$ is a $r\times r$ diagonal matrix consisting of the $r$ largest eigenvalues of $\wh\bG'\wh\bG/(k_mT)$. By (\ref{constr}), Assumptions \ref{asm1} and \ref{asm3}, we can show that $\|\wh\bV_g\|_2=O_p(1)$ and $\|\wh\bV_g^{-1}\|_2=O_p(1)$. Note that
\begin{align}\label{f:dif}
\wh\bff_t-\bK'\bff_t=&\left\{\wh\bff_t-\wh\bV_g^{-1}\left(\frac{\wh\bF'\bG\bgg_t}{Tk_m}\right)\right\}+\left\{\wh\bV_g^{-1}\left[\frac{\wh\bF'\bF\bB'\bu_t}{Tk_m}+\frac{\wh\bF'\bU\bB\bff_t}{Tk_m}+\frac{\wh\bF'\bU\bu_t}{Tk_m}\right]\right\}\notag\\
=:&\, \ba_t+\bb_t,
\end{align}
where $\bG=(\bG_1,...,\bG_m)$ and $\bU=(\bu_1,...,\bu_T)'$.
By the proof of Theorem 1 in \cite{BaiNg_Econometrica_2002} and Lemma A.1 in \cite{Bai_Econometrica_2003},
\begin{equation}\label{sum:bt}
\frac{1}{T}\sum_{t=1}^T\|\bb_t\|_2^2=O_p\left(\frac{1}{k_m}+\frac{1}{T}\right).
\end{equation}
On the other hand, note that $\wh\bff_t-\wh\bV_g^{-1}\left\{\frac{\wh\bF'\wh\bG\wh\bgg_t}{Tk_m}\right\}=0$. Therefore,
\begin{align}\label{at}
\ba_t=&\wh\bff_t-\wh\bV_g^{-1}\left\{\frac{\wh\bF'(\wh\bG\bH'+\bG-\wh\bG\bH')(\bH\wh\bgg_t+\bgg_t-\bH\wh\bgg_t)}{Tk_m}\right\}\notag\\
=&-\wh\bV_g^{-1}\left\{\frac{\wh\bF'\wh\bG(\bH'\bH-\bI_{k_m})\wh\bgg_t}{Tk_m}+\frac{\wh\bF'\wh\bG\bH'(\bgg_t-\bH\wh\bgg_t)}{Tk_m}+\frac{\wh\bF'(\bG-\wh\bG\bH')\bH\wh\bgg_t}{Tk_m}\right.\notag\\
&\left.+\frac{\wh\bF'(\bG-\wh\bG\bH')(\bgg_t-\bH\wh\bgg_t)}{Tk_m}\right\}\notag\\
=:&-\wh\bV_g^{-1}\{\Pi_{1t}+\Pi_{2t}+\Pi_{3t}+\Pi_{4t}\}.
\end{align}
By the proofs in \cite{BaiNg_joe_2006} and the Bonferroni's inequality,
\begin{equation}\label{H:rate}
\|\bH'\bH-\bI_{k_m}\|_2\leq \max_{1\leq i\leq m}\|\bH_i'\bH_i-\bI_{r_i}\|_2\leq O_p\left(\frac{m}{T}+\frac{m}{{\underline{p}}}\right),
\end{equation}
and the same result holds for $\|\bH\bH'-\bI_{k_m}\|_2$. Therefore,
\begin{align}\label{pi1}
\frac{1}{T}\sum_{t=1}^T\|\Pi_{1t}\|_2^2\leq &\|\frac{1}{T}\sum_{t=1}^T\wh\bff_t\left(\frac{\wh\bgg_t}{\sqrt{k_m}}\right)'\|_2^2\|\bH'\bH-\bI_{k_m}\|_2^2\frac{1}{T}\sum_{t=1}^T\|\frac{\wh\bgg_t}{\sqrt{k_m}}\|_2^2\leq O_p\left(\frac{m^2}{{T^2}}+\frac{m^2}{{\underline{p}^2}}\right).
\end{align}
\begin{align}\label{pi2}
\frac{1}{T}\sum_{t=1}^T\|\Pi_{2t}\|_2^2\leq &\|\frac{1}{T}\sum_{t=1}^T\wh\bff_t\left(\frac{\wh\bgg_t}{\sqrt{k_m}}\right)'\|_2^2[\frac{1}{T}\sum_{t=1}^T\|\frac{\bH'\bgg_t-\wh\bgg_t}{\sqrt{k_m}}\|_2^2+\|\bH'\bH-\bI_{k_m}\|_2^2\frac{1}{T}\sum_{t=1}^T\|\frac{\wh\bgg_t}{\sqrt{k_m}}\|_2^2]\notag\\
\leq &O_p\left\{\frac{m}{k_m}\left(\frac{1}{T}+\frac{1}{\underline{p}}\right)+\frac{m^2}{{T^2}}+\frac{m^2}{{\underline{p}^2}}\right\},
\end{align}
\begin{align}\label{pi3}
\frac{1}{T}\sum_{t=1}^T\|\Pi_{3t}\|_2^2\leq &\left[\|\frac{1}{T}\sum_{t=1}^T\wh\bff_t\left(\frac{\bH'\bgg_t-\wh\bgg_t}{\sqrt{k_m}}\right)'\|_2^2+\|\frac{1}{T}\sum_{t=1}^T\wh\bff_t(\frac{\bgg_t}{\sqrt{k_m}})'(\bI_{k_m}-\bH'\bH)'\|_2^2\right]\frac{1}{T}\sum_{t=1}^T\|\frac{\wh\bgg_t}{\sqrt{k_m}}\|_2^2\notag\\
\leq&O_p\left\{\frac{m}{k_m}\left(\frac{1}{T}+\frac{1}{\underline{p}}\right)+\frac{m^2}{{T^2}}+\frac{m^2}{{\underline{p}^2}}\right\}.
\end{align}
By a similar argument, we can show that
\begin{equation}\label{pi4}
\frac{1}{T}\sum_{t=1}^T\|\Pi_{4t}\|_2^2\leq C(\frac{1}{T}\sum_{t=1}^T\|\Pi_{2t}\|_2^2+\frac{1}{T}\sum_{t=1}^T\|\Pi_{3t}\|_2^2)=O_p\left\{\frac{m}{k_m}\left(\frac{1}{T}+\frac{1}{\underline{p}}\right)+\frac{m^2}{{T^2}}+\frac{m^2}{{\underline{p}^2}}\right\}.
\end{equation}
By (\ref{f:dif}), (\ref{sum:bt})--(\ref{pi4}), we have 
\begin{equation}\label{f:rate}
\frac{1}{T}\sum_{t=1}^T\|\wh\bff_t-\bK'\wh\bff_t\|_2^2=O_p\left(\frac{1}{k_m}+\frac{1}{T}+\frac{m}{k_m\underline{p}}+\frac{m^2}{{T^2}}+\frac{m^2}{{\underline{p}^2}}\right).
\end{equation}
(\ref{rate:f}) follows from the above inequality.\\
To prove (\ref{max:rate2}), by the proof of Theorem 4 of \cite{fan2013}, we have
\begin{equation}\label{max:bt}
\max_{1\leq t\leq T}\|\bb_t\|_2=O_p\left(\frac{1}{T}+\frac{T^{1/4}}{\sqrt{k_m}}\right).
\end{equation}
Furthermore, if $T^{1/4}=o(\underline{p}^{1/2})$, by Assumption \ref{asm2}, $\max_{t\leq T}\|\bgg_{i,t}\|_2=O_p(\log(T))$, and
\begin{align}\label{pi1:max}
\max_{t\leq T}\|\Pi_{1t}\|_2\leq &\|\frac{\wh\bF'\wh\bG}{T\sqrt{k_m}}\|_2\|\bH'\bH-\bI_{k_m}\|_2\max_{t\leq T}\left(\|\frac{\wh\bgg_t-\bH'\bgg_t}{\sqrt{k_m}}\|_2+\|\frac{\bH'\bgg_t}{\sqrt{k_m}}\|_2\right)\notag\\
\leq &O_p(1)O_p\left({\frac{m}{{T}}}+{\frac{m}{{\underline{p}}}}\right)O_p\left(\sqrt{\frac{m}{k_m}}\left(\frac{1}{T}+\frac{T^{1/4}}{\sqrt{\underline{p}}}\right)+\sqrt{\frac{m}{k_m}}\log(T)\right)\notag\\
\leq&O_p\left\{\sqrt{\frac{m}{k_m}}\log(T)\left({\frac{m}{{T}}}+{\frac{m}{{\underline{p}}}}\right)\right\}.
\end{align}
\begin{align}\label{pi2:max}
\max_{t\leq T}\|\Pi_{2t}\|_2\leq &\|\frac{\wh\bF'\wh\bG}{T\sqrt{k_m}}\|_2\max_{t\leq T}\left(\|\frac{\bH'\bgg_t-\wh\bgg_t}{\sqrt{k_m}}\|_2+\|\bI_{k_m}-\bH'\bH\|_2\|\frac{\wh\bgg_t}{\sqrt{k_m}}\|_2\right)\notag\\
\leq&O_p(1)O_p\left\{\sqrt{\frac{m}{k_m}}\left(\frac{1}{T}+\frac{T^{1/4}}{\sqrt{\underline{p}}}\right)+\sqrt{\frac{m}{k_m}}\log(T)\left({\frac{m}{{T}}}+{\frac{m}{{\underline{p}}}}\right)\right\}\notag\\
\leq &O_p\left\{\sqrt{\frac{m}{k_m}}\left[\frac{T^{1/4}}{\sqrt{\underline{p}}}+\log(T)\left(\frac{m}{T}+\frac{m}{\underline{p}}\right)\right]\right\}.
\end{align}
\begin{align}\label{pi3:max}
\max_{t\leq T}\|\Pi_{3t}\|_2\leq &\|\frac{\wh\bF'(\bG-\wh\bG\bH')\bH}{T\sqrt{k_m}}\|_2\max_{t\leq T}\|\frac{\wh\bgg_t}{\sqrt{k_m}}\|_2\notag\\
\leq & \left[\|\frac{\wh\bF'(\bG\bH-\wh\bG)}{T\sqrt{k_m}}\|_2+\|\frac{\wh\bF'\wh\bG(\bI_{k_m}-\bH'\bH)}{T\sqrt{k_m}}\|_2\right]\max_{t\leq T}\|\frac{\wh\bgg_t}{\sqrt{k_m}}\|_2\notag\\
\leq &O_p\left(\frac{1}{\sqrt{T}}+\frac{1}{\sqrt{\underline{p}}}+\frac{m}{T}+\frac{m}{\underline{p}}\right)O_p\left(\sqrt{\frac{m}{k_m}}\log(T)\right)\notag\\
\leq &O_p\left\{\sqrt{\frac{m}{k_m}}\log(T)\left[\frac{1}{\sqrt{T}}+\frac{1}{\sqrt{\underline{p}}}+\frac{m}{T}+\frac{m}{\underline{p}}\right]\right\},
\end{align}
and $\max_{t\leq T}\|\Pi_{4t}\|_2\leq \max_{t\leq T}\|\Pi_{3t}\|_2$ implies that
\begin{equation}\label{pi4:max}
\max_{t\leq T}\|\Pi_{4t}\|_2=O_p\left\{\sqrt{\frac{m}{k_m}}\log(T)\left[\frac{1}{\sqrt{T}}+\frac{1}{\sqrt{\underline{p}}}+\frac{m}{T}+\frac{m}{\underline{p}}\right]\right\}.
\end{equation}
(\ref{max:rate2}) follows from Equations (\ref{max:bt})--(\ref{pi4:max}). This completes the proof of Theorem 1. $\Box$

{\bf Proof of Theorem 2.} We only prove  the result of the second one, as the first one follows from Theorem 4 of \cite{fan2013}. 

Let $\bC_i=\bA_i\bB_i$ and $\wh\bC_i=\wh\bA_i\wh\bB_i$. We note that $\wh\bA_i=\bY_i'\wh\bG_i/T$ and $\wh\bB_i=\wh\bG_i'\wh\bF/T$, then for $1\leq l\leq p_i$,
\begin{equation}\label{cil}
\wh\bfc_{i,l}=\frac{1}{T}\sum_{t=1}^T\wh\bff_t\wh\bgg_{i,t}'\frac{1}{T}\sum_{t=1}^T\wh\bgg_{i,t}y_{i,lt}.
\end{equation}
Define
\[S_{1t}=\frac{1}{T}\sum_{t=1}^T(\wh\bff_t-\bK'\bff_t)\wh\bgg_{i,t}'\frac{1}{T}\sum_{t=1}^T\wh\bgg_{i,t}y_{i,lt},\,\, S_{2t}=\bK'\frac{1}{T}\sum_{t=1}^T\bff_t(\wh\bgg_{i,t}-\bH_i'\bgg_{i,t})'\frac{1}{T}\sum_{t=1}^T\wh\bgg_{i,t}y_{i,lt},\]
\[S_{3t}=\bK'\frac{1}{T}\sum_{t=1}^T\bff_t\bgg_{i,t}'\bH_i\frac{1}{T}\sum_{t=1}^T(\wh\bgg_{i,t}-\bH_i'\bgg_{i,t})y_{i,lt},\,\,S_{4t}=\bK'\frac{1}{T}\sum_{t=1}^T\bff_t\bgg_{i,t}'(\bH_i\bH_i'-\bI_{r_i})\frac{1}{T}\sum_{t=1}^T\bgg_{i,t}y_{i,lt},\]
and
\[S_{5t}=\bK'\frac{1}{T}\sum_{t=1}^T\bff_t\bgg_{i,t}'\frac{1}{T}\sum_{t=1}^T\bgg_{i,t}u_{i,lt}.\]
It follows from (\ref{cil}) that
\begin{equation}\label{cil:dif}
\wh\bfc_{i,l}-\bK'\bfc_{i,l}=S_{1t}+S_{2t}+S_{3t}+S_{4t}+S_{5t}.
\end{equation}
Note that
\[\frac{1}{T}\sum_{t=1}^T\|\wh\bff_t-\bK'\bff_t\|_2^2=O_p\left(\frac{1}{k_m}+\frac{1}{T}+\frac{m}{k_m\underline{p}}+\frac{m^2}{{T^2}}+\frac{m^2}{{\underline{p}^2}}\right),\]
\[\,\,\frac{1}{T}\sum_{t=1}^T\|\wh\bgg_t\|_2^2=O_p(1),\,\,\|\bH_i\bH_i'-\bI_{r_i}\|_2^2=O_p\left(\frac{1}{T^2}+\frac{1}{\underline{p}^2}\right),\]
and $\max_{l\leq p_i}\frac{1}{T}\sum_{t=1}^Ty_{i,lt}^2=O_p(1)$, then, by the Cauchy-Schwarz inequality,
\begin{align}\label{s1t}
\max_{l\leq p_i}\|S_{1t}\|_2\leq& (\frac{1}{T}\sum_{t=1}^T\|\wh\bff_t-\bK'\bff_t\|_2^2\frac{1}{T}\sum_{t=1}^T\|\wh\bgg_{i,t}\|_2^2)^{1/2}(\frac{1}{T}\sum_{t=1}^T\|\wh\bgg_{i,t}\|_2^2\max_{l\leq p_i}\frac{1}{T}\sum_{t=1}^Ty_{i,lt}^2)^{1/2}\notag\\
=&O_p\left(\sqrt{\frac{1}{k_m}+\frac{1}{T}+\frac{m}{k_m\underline{p}}+\frac{m^2}{{T^2}}+\frac{m^2}{{\underline{p}^2}}}\right).
\end{align}
Similarly,
\[\max_{l\leq p_i}\|S_{2t}\|_2=O_p\left(\frac{1}{\sqrt{T}}+\frac{1}{\sqrt{\underline{p}}}\right),\,\,\max_{l\leq p_i}\|S_{3t}\|_2=O_p\left(\frac{1}{\sqrt{T}}+\frac{1}{\sqrt{\underline{p}}}\right),\]
and
\[\max_{l\leq p_i}\|S_{4t}\|_2=O_p\left(\frac{1}{{T}}+\frac{1}{{\underline{p}}}\right).\]
By Assumption \ref{asm3},
\[\max_{l\leq p_i}\|S_{5t}\|_2=O_p\left(\sqrt{\frac{\log(p_i)}{T}}\right)\]
Therefore,
\[\max_{l\leq p_i}\|\wh\bfc_{i,l}-\bK'\bfc_{i,l}\|_2=O_p\left(\sqrt{\frac{1}{k_m}+\frac{m}{k_m\underline{p}}+\frac{m^2}{{T^2}}+\frac{m^2}{{\underline{p}^2}}}+\sqrt{\frac{\log(p_i)}{T}}\right).\]
Furthermore, note that
\begin{align}\label{gt:df}
\wh\ba_{i,l}'\wh\bgg_{i,t}-\ba_{i,l}'\bgg_{i,t}=&(\wh\ba_{i,l}-\bH_i'\ba_{i,l})'(\wh\bgg_{i,t}-\bH_i'\bgg_{i,t})+(\wh\ba_{i,l}-\bH_i'\ba_{i,l})'\bH_i'\bgg_{i,t}\notag\\
&+(\bH_i'\ba_{i,l})'(\wh\bgg_{i,t}-\bH_i'\bgg_{i,t})+\ba_{i,l}'(\bH_i\bH_i'-\bI_{r_i})\bgg_{i,t}.
\end{align}
Therefore,
\begin{align}
\|\wh\ba_{i,l}'\wh\bgg_{i,t}-\ba_{i,l}'\bgg_{i,t}\|_2\leq &\|\wh\ba_{i,l}-\bH_i'\ba_{i,l}\|_2\|\wh\bgg_{i,t}-\bH_i'\bgg_{i,t}\|_2+\|\wh\ba_{i,l}-\bH_i'\ba_{i,l}\|_2\|\bH_i'\bgg_{i,t}\|_2\notag\\
&+\|\bH_i'\ba_{i,l}\|_2\|\wh\bgg_{i,t}-\bH_i'\bgg_{i,t}\|_2+\|\ba_{i,l}\|_2\|(\bH_i\bH_i'-\bI_{r_i}\|_2\|\bgg_{i,t}\|_2.
\end{align}
By Assumption \ref{asm3} and the Bonferroni's inequality, $\max_{t\leq T}\|\bgg_{i,t}\|_2=O_p(\log(T))$. It follows from (\ref{max:rate1}) and (\ref{max-loading1}) that
\[\max_{l\leq p_i,t\leq T}\|\wh\ba_{i,l}'\wh\bgg_{i,t}-\ba_{i,l}'\bgg_{i,t}\|_2=O_p\left(\log(T)\sqrt{\frac{\log(p_i)}{T}}+\frac{T^{1/4}}{\sqrt{p_i}}\right).\]
Similarly, we can show that 
\[\max_{l\leq p_i,t\leq T}\|\wh\bfc_{i,l}'\wh\bff_t-\bfc_{i,l}'\bff_t\|_2=O_p\left(\log(T)w_T+\sqrt{\frac{m}{k_m}}\frac{T^{1/4}}{\sqrt{\underline{p}}}+\frac{T^{1/4}}{\sqrt{k_m}}\right),\]
where $w_{T}=\sqrt{\frac{m}{k_m\underline{p}}+\frac{m^2}{{T^2}}+\frac{m^2}{{\underline{p}^2}}}+\sqrt{\frac{\log(p_i)}{T}}$.
This completes the proof of Theorem 2. $\Box$

{\bf Proof of Theorem 3.} We temporarily impose the identification conditions that $\bA_i'\bA_i=\bI_{r_i}$ and $\bB'\bB=\bI_r$. By (\ref{m-factor}),
\[\bSigma_{y_i}=\bA_i\bSigma_{g_i}\bA_i'+\bSigma_{e_i},\]
where $\|\bSigma_{g_i}\|_{\min}\asymp\|\bSigma_{g_i}\|_2\asymp p_i$ according to Assumption \ref{asm1}. Note that
\[\wh\bSigma_{y_i}=\bA_i\wh\bSigma_{g_i}\bA_i'+\bA_i\wh\bSigma_{g_ie_i}+\wh\bSigma_{e_ig_i}\bA_i'+\wh\bSigma_{e_i.}\]
By Assumptions \ref{asm2}--\ref{asm3} and the proofs in \cite{gaotsay2020b}, we have
\[\|\wh\bSigma_{g_i}-\bSigma_{g_i}\|_2=O_p(p_iT^{-1/2}),\,\,\|\bA_i\wh\bSigma_{g_ie_i}\|_2=O_p(p_iT^{-1/2}),\]
\[\|\wh\bSigma_{e_ig_i}\|_2=O_p(p_iT^{-1/2}),\,\,\|\wh\bSigma_{e_i}-\bSigma_{e_i}\|_2=O_p(p_iT^{-1/2}).\]
By Lemma 1 in \cite{gaotsay2020b} and $\lambda_{r_i}(\bSigma_{y_i})\geq Cp_i$,
\[\|\wh\bA_i-\bA_i\|_2\leq \frac{\|\wh\bSigma_{y_i}-\bSigma_{y_i}\|_2}{\lambda_{r_i}(\bSigma_{y_i})}=O_p(T^{-1/2}).\]
Note that $\wh\bgg_{i,t}=\wh\bA_i'\by_{i,t}=\wh\bA_i'\bA_i\bgg_{i,t}+\wh\bA_i'\be_{i,t}$, then
\begin{align}\label{ag:d}
\|\wh\bA_i\wh\bgg_{i,t}-\bA_i\bgg_{i,t}\|_2\leq&\|(\wh\bA_i-\bA_i)\bgg_{i,t}\|_2+\|\wh\bA_i\wh\bA_i'(\bA_i-\wh\bA_i)\bgg_{i,t}\|_2+\|\wh\bA_i\wh\bA_i'\be_{i,t}\|_2\notag\\
\leq &O_p(p_i^{1/2}T^{-1/2})+O_p(p_i^{1/2}T^{-1/2})+O_p(1),
\end{align}
where the last term follows from the fact that the dimension of $\wh\bA_i'\be_{i,t}$ is finite and its variance is also finite. Therefore,
\[p_i^{-1/2}\|\wh\bA_i\wh\bgg_{i,t}-\bA_i\bgg_{i,t}\|_2=O_p(p_i^{-1/2}+T^{-1/2}).\]
Next, we observe that $\wh\bgg_{t}=\wh\bA'\bA\bgg_{t}+\wh\bA'\be_{t}$, and
\begin{align}\label{sgi:t}
\wt\bSigma_{g}:=\frac{1}{T}\sum_{t=1}^T\wh\bgg_{t}\wh\bgg_{t}'=&\wh\bA'\bA\wh\bSigma_{g}\bA'\wh\bA+\wh\bA'\wh\bSigma_{e}\wh\bA+\wh\bA'\bA\wh\bSigma_{ge}\wh\bA+\wh\bA'\wh\bSigma_{eg}\bA\wh\bA\notag\\
=&\wh\bSigma_{g}+(\wh\bA-\bA)'\bA\wh\bSigma_{g}+\wh\bSigma_{g}\bA'(\wh\bA-\bA)+(\wh\bA-\bA)'\bA\wh\bSigma_{g}\bA'(\wh\bA-\bA)\notag\\
&+\wh\bA'\wh\bSigma_{e}\wh\bA+\wh\bA'\bA\wh\bSigma_{ge}\wh\bA+\wh\bA'\wh\bSigma_{eg}\bA'\wh\bA\notag\\
=:&J_1+J_2+...+J_7.
\end{align}
By a similar argument as above and the proofs in \cite{gaotsay2020b},
\[\|J_1-\bSigma_{g}\|_2=O_p(k_m\bar{p}T^{-1/2}),\,\,\|J_2\|_2=O_p(\bar{p}k_mT^{-1/2}),\,\, \|J_3\|_2=O_p(\bar{p}k_mT^{-1/2})\]
\[\|J_4\|_2=O_p(\bar{p}k_mT^{-1}),\,\,\|J_5\|_2=O_p(k_m),\,\,\|J_6\|_2=O_p(k_m\bar{p}^{1/2}T^{-1/2}),\,\,\|J_7\|_2=O_p(k_m\bar{p}^{1/2}T^{-1/2}).\]
Note that $\lambda_{r}(\bSigma_g)\geq \underline{p}k_m$, by Lemma 1 in \cite{gaotsay2020b} again,
\[\|\wh\bB-\bB\|_2\leq \frac{\|\wt\bSigma_g-\bSigma_g\|_2}{ \underline{p}k_m}=O_p(\underline{p}^{-1}\bar{p}T^{-1/2}+\underline{p}^{-1}),\]
which implies that $\|\wh\bB_i-\bB_i\|_2=O_p(\underline{p}^{-1}\bar{p}T^{-1/2}+\underline{p}^{-1})$.
Then,
\[\|\wh\bA_i\wh\bB_i-\bA_i\bB_i\|_2=O_p(\underline{p}^{-1}\bar{p}T^{-1/2}+\underline{p}^{-1})\]
and 
\[\|\wh\bA\wh\bB-\bA\bB\|_2\leq \|(\wh\bA-\bA)\wh\bB\|_2+\|\bA(\wh\bB-\bB)\|_2=O_p(mT^{-1/2}+\underline{p}^{-1}\bar{p}T^{-1/2}+\underline{p}^{-1}).\]
In addition, as $\wh\bff_t=\wh\bB'\wh\bA'\by_t$, and
\begin{align}\label{abf:d}
\wh\bA\wh\bB\wh\bff_t-\bA\bB\bff_t=&(\wh\bA\wh\bB-\bA\bB)\bff_t+\wh\bA\wh\bB\wh\bB'\wh\bA(\bA\bB-\wh\bA\wh\bB)\bff_t+\wh\bA\wh\bB\wh\bB'\wh\bA'\bA\bu_t+\wh\bA\wh\bB\wh\bB'\wh\bA'\be_t.
\end{align}
Let $\delta_T=mT^{-1/2}+\underline{p}^{-1}\bar{p}T^{-1/2}+\underline{p}^{-1}$, by a similar argument as the proof of Theorem 4 in \cite{gaotsay2020b}, we have
\[\|(\wh\bA\wh\bB-\bA\bB)\bff_t\|_2=O_p(k_m^{1/2}\delta_T),\,\,\|\wh\bA\wh\bB\wh\bB'\wh\bA'(\bA\bB-\wh\bA\wh\bB)\bff_t\|_2=O_p(k_m^{1/2}\delta_T),\]
\[\|\wh\bA\wh\bB\wh\bB'\wh\bA'\bA\bu_t\|_2\leq\|\wh\bB'\wh\bA'\bA\bu_t\|_2=O_p(1), \]
and
\[\|\wh\bA\wh\bB\wh\bB'\wh\bA'\be_t\|_2\leq \|\wh\bB'\wh\bA'\be_t\|_2=O_p(1).\]
Therefore,
\[p^{-1/2}\|\wh\bA\wh\bB\wh\bff_t-\bA\bB\bff_t\|_2=O_p(p^{-1/2}k_m^{1/2}\delta_T+p^{-1/2}).\]
This completes the proof. $\Box$

{\bf Proof of Theorem 4.} The consistency of $\wh r_i$ follows from Theorem 2 of \cite{BaiNg_Econometrica_2002}. We only show the consistency of the ratio-based method. Let $\lambda_1\geq...\geq \lambda_{k_m}$ be the top $k_m$ eigenvalues of $\bG\bG'$.
Note that
\begin{align}\label{GGT}
\wh\bG\wh\bG'=&\bG\bH\bH'\bG+\bG\bH(\wh\bG-\bG\bH)'+(\wh\bG-\bG\bH)\bH'\bG'+(\wh\bG-\bG\bH)(\wh\bG-\bG\bH)'\notag\\
=&\bG\bG'+\bG(\bH\bH'-\bI_{k_m})\bG+\bG\bH(\wh\bG-\bG\bH)'+(\wh\bG-\bG\bH)\bH'\bG'\notag\\
&+(\wh\bG-\bG\bH)(\wh\bG-\bG\bH)'.
\end{align}
By Assumption \ref{asm1}--\ref{asm3}, we have that $\lambda_1(\bG\bG'/(k_mT))\asymp...\asymp \lambda_{r}(\bG\bG'/(k_mT))\asymp 1$, and $\lambda_{r+j}(\bG\bG'/(k_mT))\asymp k_m^{-1}$ for $1\leq j\leq \min(k_m-r, T-r)$. By Weyl's inequality and the previous arguments, we have
\[|\wh\lambda_j(\wh\bG\wh\bG'/(k_mT))-\lambda_j(\bG\bG'/(k_mT))|=o_p(1),\] 
for $1\leq j\leq \min(k_m,T)$. Therefore, $\wh\lambda_j(\wh\bG\wh\bG)\asymp k_mT$ for $1\leq j\leq r$ and 
$\wh\lambda_j(\wh\bG\wh\bG)\asymp T$ for $r+1\leq j\leq \min(k_m,T)$. Therefore, there is a sharp drop at $l=r$ in (\ref{est-r}). This completes the proof of Theorem 3. $\Box$

{\bf Proof of Theorem 5.}  First, note that
\[\wh\bSigma_{y_i}=\bA_i\wh\bSigma_{g_i}\bA_i'+\bA_i\wh\bSigma_{g_ie_i}+\wh\bSigma_{e_ig_i}\bA_i'+\wh\bSigma_{e_i},\]
and by Assumptions \ref{asm1}, \ref{asm3}(i), \ref{asm4}(i)-(iv), \ref{asm5}-\ref{asm6}, 
\[\|\wh\bSigma_{y_i}-\bA_i\wh\bSigma_{g_i}\bA_i'\|_2\leq C\|\wh\bSigma_{g_ie_i}\|_2+C\|\wh\bSigma_{e_i}\|_2\leq Cp_i^{1/2}\sqrt{o_p(p_iT^{2\delta})}+O_p(1)=o_p(p_iT^{\delta}).\]
Since $\lambda_{r_i}(\bA_i\wh\bSigma_{g_i}\bA_i')\geq Cp_iT$, by Lemma 1 in \cite{gaotsay2020b},
\[\|\wh\bA_i-\bA_i\|_2\leq C\frac{\|\wh\bSigma_{y_i}-\bA_i\wh\bSigma_{g_i}\bA_i'\|_2}{\lambda_{r_i}(\bA_i\wh\bSigma_{g_i}\bA_i')}=o_p(\frac{1}{T^{1-\delta}}).\]

Next, as the strength of $\bA$ was imposed to $\bgg_{t}$, we define $\bD_p=\diag(p_i\bI_{r_1},...,p_m\bI_{r_m})$. For any given $\bA_i$,  we reformulate (\ref{global-f}) as
\[\bgg_t^*=\bB^*\bff_t+\bu_t,\]
where $\bgg_t^*=\bD_p^{-1/2}\bgg_t$ is the normalized one and $\bB^*$ is the new orthogonal matrix that we are going to estimate, and $\bff_t$ has an additional strength of $k_m^{1/2}$. We still use $\bu_t$ because it only differs from the one in (\ref{global-f}) up to an orthogonal transformation.
 
Note that
\begin{align}\label{sig:gt}
\wt\bSigma_g=&\bD_p^{-1/2}\frac{1}{T}\sum_{t=1}^T(\wh\bA'\by_t\by_t'\wh\bA)\bD_p^{-1/2}\notag\\
=&(\bD_p^{-1/2}\wh\bA'\bA)\frac{1}{T}\sum_{t=1}^T\bgg_t\bgg_t'(\bA'\wh\bA\bD_p^{-1/2})+(\bD_p^{-1/2}\wh\bA'\bA)\frac{1}{T}\sum_{t=1}^T\bgg_t\be_t'(\bA'\wh\bA\bD_p^{-1/2})\notag\\
&+(\bD_p^{-1/2}\wh\bA'\bA)\frac{1}{T}\sum_{t=1}^T\be_t\bgg_t'(\bA'\wh\bA\bD_p^{-1/2})+(\bD_p^{-1/2}\wh\bA'\bA)\frac{1}{T}\sum_{t=1}^T\be_t\be_t'(\bA'\wh\bA\bD_p^{-1/2})\notag\\
=:&R_{1}+R_2+R_3+R_4.
\end{align}
 We now compute all the convergence rates of the above terms. By the assumptions in Theorem 5,
 \begin{align}\label{r2t}
 \|R_2\|_2=C\frac{1}{\underline{p}}\|\frac{1}{T}\sum_{t=1}^T(\bB\bff_t+\bu_t)\be_t\|_2\leq& C\frac{1}{\underline{p}}\left(\|\frac{1}{T}\sum_{t=1}^T\bB\bff_t\be_t\|_2+\|\frac{1}{T}\sum_{t=1}^T\bu_t\be_t\|_2\right)\notag\\
 \leq &C\frac{1}{\underline{p}} \left(k_m^{1/2}\bar{p}^{1/2}O_p(\sqrt{pT^{2\delta}})+O_p(\bar{p}^{1/2}\sqrt{k_mpT^{-1}})\right)\notag\\
 \leq&O_p(\underline{p}^{-1}k_m^{1/2}\bar{p}^{1/2}p^{1/2}T^\delta),
 \end{align}
and $\|R_3\|_2=\|R_2\|_2=O_p(\underline{p}^{-1}k_m^{1/2}\bar{p}^{1/2}p^{1/2}T^\delta)$. 
\begin{equation}\label{r4t}
\|R_{4}\|_2\leq C\frac{1}{\underline{p}}\|\frac{1}{T}\sum_{t=1}^T\be_t\be_t')\|_2=O_p(\underline{p}^{-1}k_m).
\end{equation}
\begin{align}\label{r1t}
R_1=&(\bD_p^{-1/2}\wh\bA'\bA)\frac{1}{T}\sum_{t=1}^T\bB\bff_t\bff_t'\bB{'}(\bA'\wh\bA\bD_p^{-1/2})+(\bD_p^{-1/2}\wh\bA'\bA)\frac{1}{T}\sum_{t=1}^T\bB\bff_t\bu_t'(\bA'\wh\bA\bD_p^{-1/2})\notag\\
&+(\bD_p^{-1/2}\wh\bA'\bA)\frac{1}{T}\sum_{t=1}^T\bu_t\bff_t'\bB{'}(\bA'\wh\bA\bD_p^{-1/2})+(\bD_p^{-1/2}\wh\bA'\bA)\frac{1}{T}\sum_{t=1}^T\bu_t\bu_t'(\bA'\wh\bA\bD_p^{-1/2})\notag\\
=:&\Delta_1+\Delta_2+\Delta_3+\Delta_4.
\end{align}
By a similar argument as above, 
\[\|\Delta_2\|_2=O_p(\underline{p}^{-1}\bar{p}k_mT^{\delta}),\|\Delta_3\|_2=O_p(\underline{p}^{-1}\bar{p}k_mT^{\delta}),\|\Delta_4\|_2=O_p(\underline{p}^{-1}\bar{p}^{1/2}).\]
As for $\Delta_1$, we note that
\begin{align}\label{delta1}
\Delta_1=&\bD_p^{-1/2}(\wh\bA-\bA)'\bA\bB(\frac{1}{T}\sum_{t=1}^T\bff_t\bff_t')\bB{'}\bA'\wh\bA\bD_p^{-1/2}+\bD_p^{-1/2}\bB(\frac{1}{T}\sum_{t=1}^T\bff_t\bff_t')\bB{'}\bA'(\wh\bA-\bA)\bD_p^{-1/2}\notag\\
&+\bD_p^{-1/2}\bB(\frac{1}{T}\sum_{t=1}^T\bff_t\bff_t')\bB{'}\bD_p^{-1/2}\notag\\
=:&\Delta_{1,1}+\Delta_{1,2}+\Delta_{1,3}.
\end{align}
Therefore,
\[\|\Delta_{1,1}\|_2\leq C\underline{p}^{-1}\bar{p}T^{\delta-1}T=O_p(\underline{p}^{-1}\bar{p}T^{\delta}),\]
and $\Delta_{1,2}\|_2=\|\Delta_{1,1}\|_2=O_p(\underline{p}^{-1}\bar{p}T^{\delta})$, and consequently,
\[\|\Delta_1-\bD_p^{-1/2}\bB(\frac{1}{T}\sum_{t=1}^T\bff_t\bff_t')\bB{'}\bD_p^{-1/2}\|_2=O_p(\underline{p}^{-1}\bar{p}T^{\delta}).\]
We combine the above rates and have
\[\|R_1-\bD_p^{-1/2}\bB(\frac{1}{T}\sum_{t=1}^T\bff_t\bff_t')\bB{'}\bD_p^{-1/2}\|_2=O_p(\underline{p}^{-1}\bar{p}^{1/2}k_mT^{\delta}),\]
and
\[\|\wt\bSigma_g-\bD_p^{-1/2}\bB(\frac{1}{T}\sum_{t=1}^T\bff_t\bff_t')\bB{'}\bD_p^{-1/2}\|_2=O_p(\underline{p}^{-1}k_m\bar{p}T^{\delta}).\]
As $\lambda_{r}(\bD_p^{-1/2}\bB(\frac{1}{T}\sum_{t=1}^T\bff_t\bff_t')\bB{'}\bD_p^{-1/2})\geq k_mT$, by Lemma 1 in \cite{gaotsay2020b},
\[\|\wh\bB^*-\bB^*\|_2=O_p\left(\frac{\bar{p}}{T^{1-\delta}\underline{p}}\right).\]
By a similar argument as that in Theorem \ref{tm3}, we have
\[\|\wh\bA\wh\bB^*-\bA\bB^*\|_2=O_p\left(\frac{m}{T^{1-\delta}}+\frac{\bar{p}}{T^{1-\delta}\underline{p}}\right).\]
 This proves (\ref{DA:DAB-1}). The rest of Theorem \ref{tm5} can be proved by a similar argument as that in Theorem \ref{tm3}. This completes the proof. $\Box$


\end{document}